\RequirePackage{fix-cm}
\documentclass[smallextended]{svjour3}       
\smartqed  
\usepackage{graphicx}
\usepackage{latexsym}
\usepackage{lscape}
\begin{document}

\title{Operations and Performance of the PACS Instrument $^3$He Sorption Cooler on board
       of the {\it Herschel} Space Observatory\thanks{{\it Herschel} is 
       an ESA space observatory with science instruments provided by 
       European-led Principal Investigator consortia and with important 
       participation from NASA.}
}

\titlerunning{PACS $^3$He Sorption Cooler Operations and Performance}

\author{M.~Sauvage \and
        K.~Okumura \and
        U.~Klaas \and
        Th.~M\"uller \and
        A.~Mo\'or \and
        A.~Poglitsch \and
        H.~Feuchtgruber \and
        L.~Duband 
}


\institute{M. Sauvage \at
           Commissariat \`a l'Energie Atomique, IRFU, Orme des Merisiers,
           B\^at.~709, 91191 Gif/Yvette, France \\
              \email{marc.sauvage@cea.fr}           
           \and
           K. Okumura \at
           Commissariat \`a l'Energie Atomique, IRFU, Orme des Merisiers,
           B\^at.~709, 91191 Gif/Yvette, France
           \and
           U. Klaas \at
           Max-Planck-Institut f\"ur Astronomie, K\"onigstuhl 17, 69117~Heidelberg, Germany 
           \and
           Th. M\"uller, A. Poglitsch, H. Feuchtgruber \at
           Max-Planck-Institut f\"ur extraterrestrische Physik, Giessenbachstra{\ss}e, 
           85748 Garching, Germany
           \and
           A. Mo\'or \at
           Konkoly Observatory, MTA CSFK, Konkoly Thege Mikl\'os \'ut 15-17, 
           1121 Budapest, Hungary
           \and
           L. Duband \at
           Commissariat \`a l'Energie Atomique, INAC, Service des Basses Temp\'eratures, 
           17, rue des Martyrs, 38054 Grenoble, France
}

\date{Received: November 16, 2013 / Accepted: April 8, 2014}

\maketitle

\begin{abstract}
A $^3$He sorption cooler produced the operational temperature
of 285\,mK for the bolometer arrays of the Photodetector Array
Camera and Spectrometer (PACS) instrument of the {\it Herschel}
Space Observatory. This cooler provided a stable hold 
time between 60 and 73\,h, depending on the operational 
conditions of the instrument. The respective hold time could be 
determined by a simple functional relation established early on 
in the mission and reliably applied by the scientific mission 
planning for the entire mission. After exhaustion of the liquid
$^3$He due to the heat input by the detector arrays, the
cooler was recycled for the next operational period following
a well established automatic procedure. We give an overview of 
the cooler operations and performance over the entire mission and 
distinguishing in-between the start conditions for the cooler 
recycling and the two main modes of PACS photometer operations.
As a spin-off, the cooler recycling temperature effects on the
{\it Herschel} cryostat $^4$He bath were utilized as an alternative
method to dedicated Direct Liquid Helium Content Measurements
in determining the lifetime of the liquid Helium coolant.
\keywords{Herschel \and PACS \and SPIRE \and telescopes \and space vehicles:
  instrumentation \and instrumentation: photometers \and calibration 
\and space cryogenics \and sorption coolers \and $^3$He system
}
\end{abstract}

\section{Introduction}
\label{sec:intro}
The {\it Herschel} Space Observatory~\cite{Refpilbratt10} with
its 3.5\,m telescope provided an excellent platform for FIR
observations with unprecedented sensitivity, photometric accuracy
and spatial resolution. The FIR instrument PACS~\cite{Refpoglitsch10}
employed the largest detector arrays~\cite{Refbillot10} ever flown 
in space. Its bolometer arrays needed to be cooled to an operational
temperature of 285\,mK, which was achieved by means of a $^3$He 
sorption cooler inside the instrument. We give a short overview
of this type of device, which was also used for the SPIRE~\cite{Refgriffin10}
instrument, the recycling process and the characterization of the 
cooler hold time, which was an essential parameter for the 
{\it Herschel} scientific mission planning. We present
a statistics and characteristics of all PACS cooler cycles over 
the entire {\it Herschel} mission. Finally we sketch a method
utilizing the thermal interaction of the sorption cooler pump 
and the $^4$He coolant in the {\it Herschel} cryostat, by dumping
a constant heat input, for determining the remaining liquid $^4$He
mass and hence its lifetime.

\section{The PACS  $^3$He Sorption Cooler}
\label{sec:cooler}

\begin{figure}[h!]
  \begin{center}
  \includegraphics[width=1.0\textwidth]{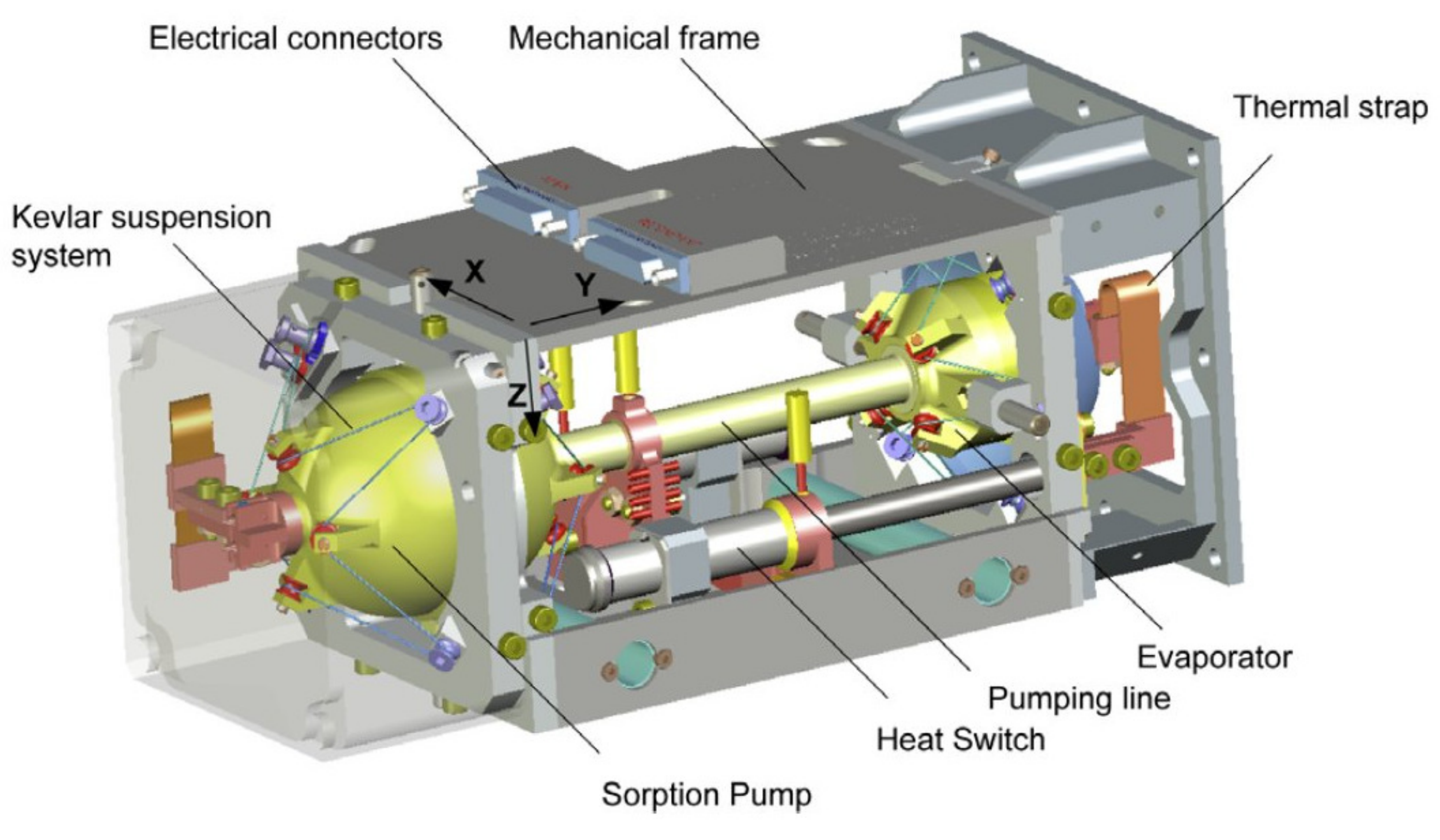}
  \includegraphics[width=1.0\textwidth]{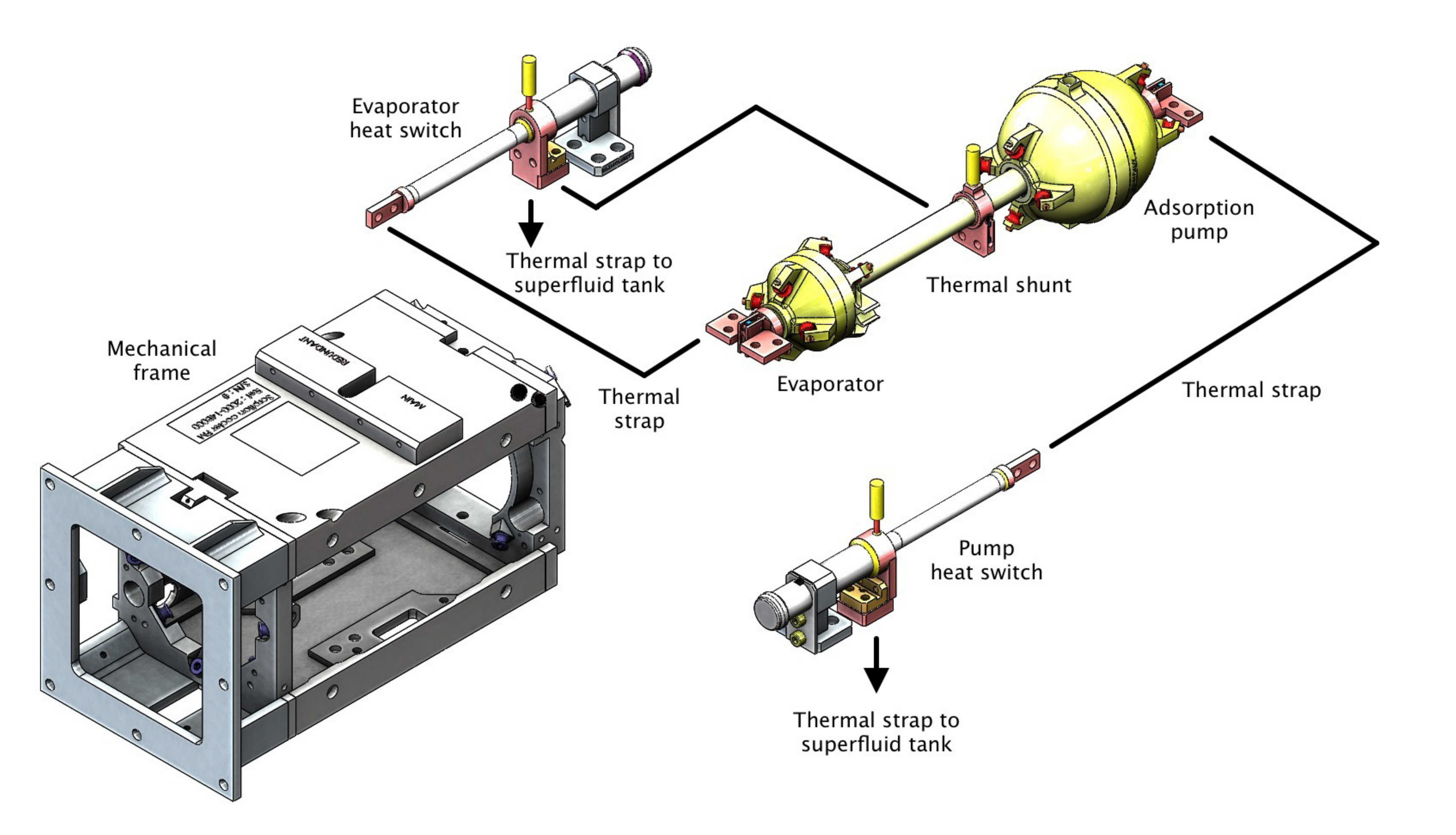}
  \end{center}
\caption{Overall 3D views of the PACS cooler identifying its main components
(from~\cite{Refduband08}). The bottom figure presents an exploded view identifying the elements represented schematically in Fig.\,\ref{fig:cooler_schematic}.
}
\label{fig:pacs_cooler}       
\end{figure}

The {\it Herschel} flight model cooler is described in detail 
in~\cite{Refduband08}. Fig.~\ref{fig:pacs_cooler} gives a 3D view
of the device showing its main components. 

The evaporator contains a porous material, an alumina sponge 
(91\,\% Al$_{2}$O$_{3}$ / 9\,\% SiO$_{2}$), which traps the liquid $^3$He 
during the cold state. This liquid $^3$He evaporates providing the 
cooling to the detector focal plane. When all the liquid $^3$He has been 
evaporated into the gas phase, it needs to be recycled.
The gaseous $^3$He flows into the sorption pump which contains 
active {\bf charcoal} for adsorption of the gas.

\section{Procedure of the Cooler Recycling}
\label{sec:cooler_recycling}

\begin{figure}[h!]
  \begin{center}
  \includegraphics[width=0.6\textwidth]{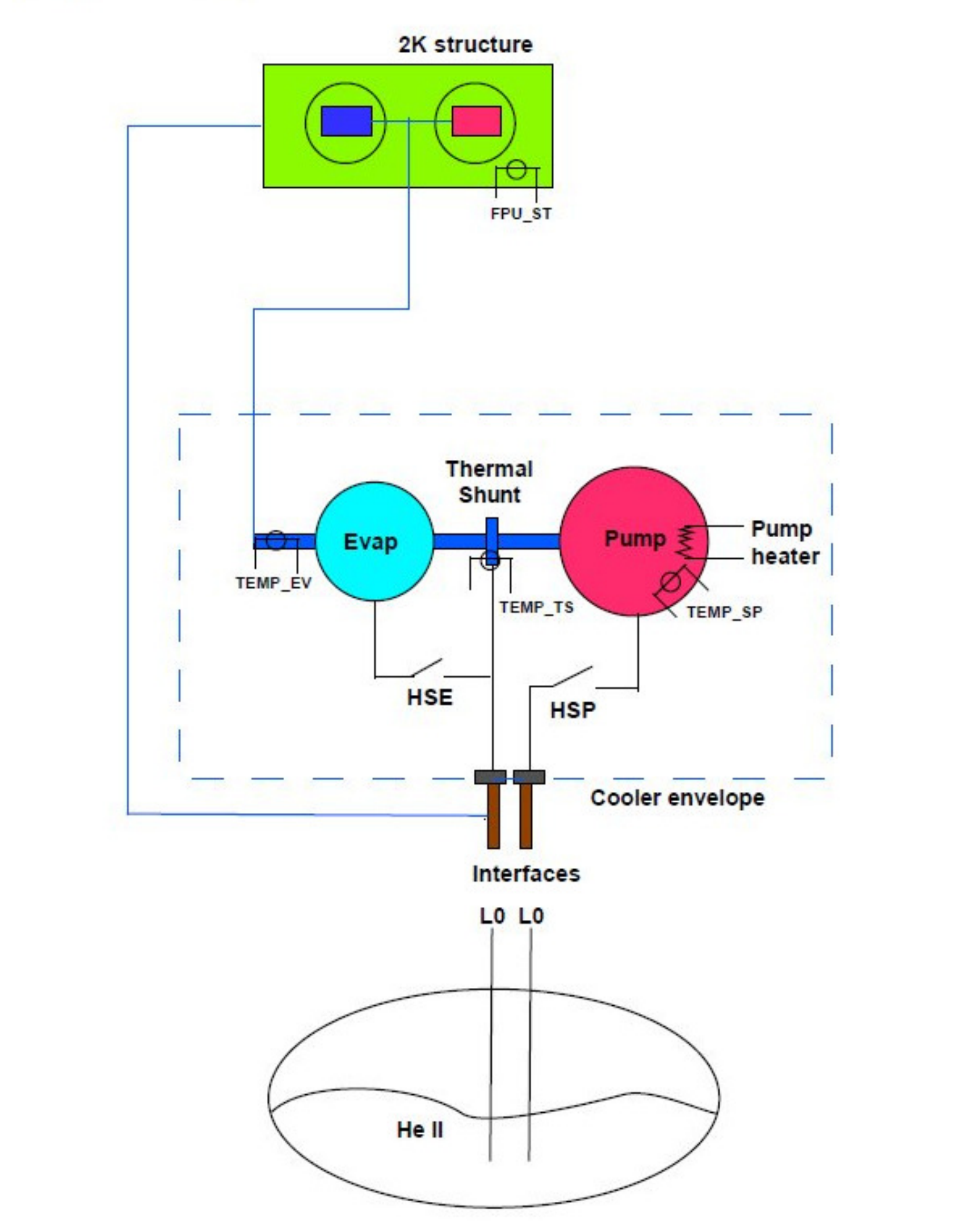}
  \end{center}
\caption{Schematic drawing of the PACS cooler elements
and the thermal connections to the PACS bolometer 
detector focal plane unit (in green) and the liquid $^4$He L0 bath of the
Herschel cryostat at $\approx$1.7\,K. 
}
\label{fig:cooler_schematic}       
\end{figure}

Fig.~\ref{fig:cooler_schematic} provides a schematic view
of the PACS cooler elements and their thermal connection
to the bolometer detector focal plane unit and the
liquid $^4$He bath of the {\it Herschel} cryostat, also referred to as the level 0 of the thermal system, L0, at $\approx$1.7\,K.

\begin{figure}[h!]
  \begin{center}
  \includegraphics[width=1.0\textwidth]{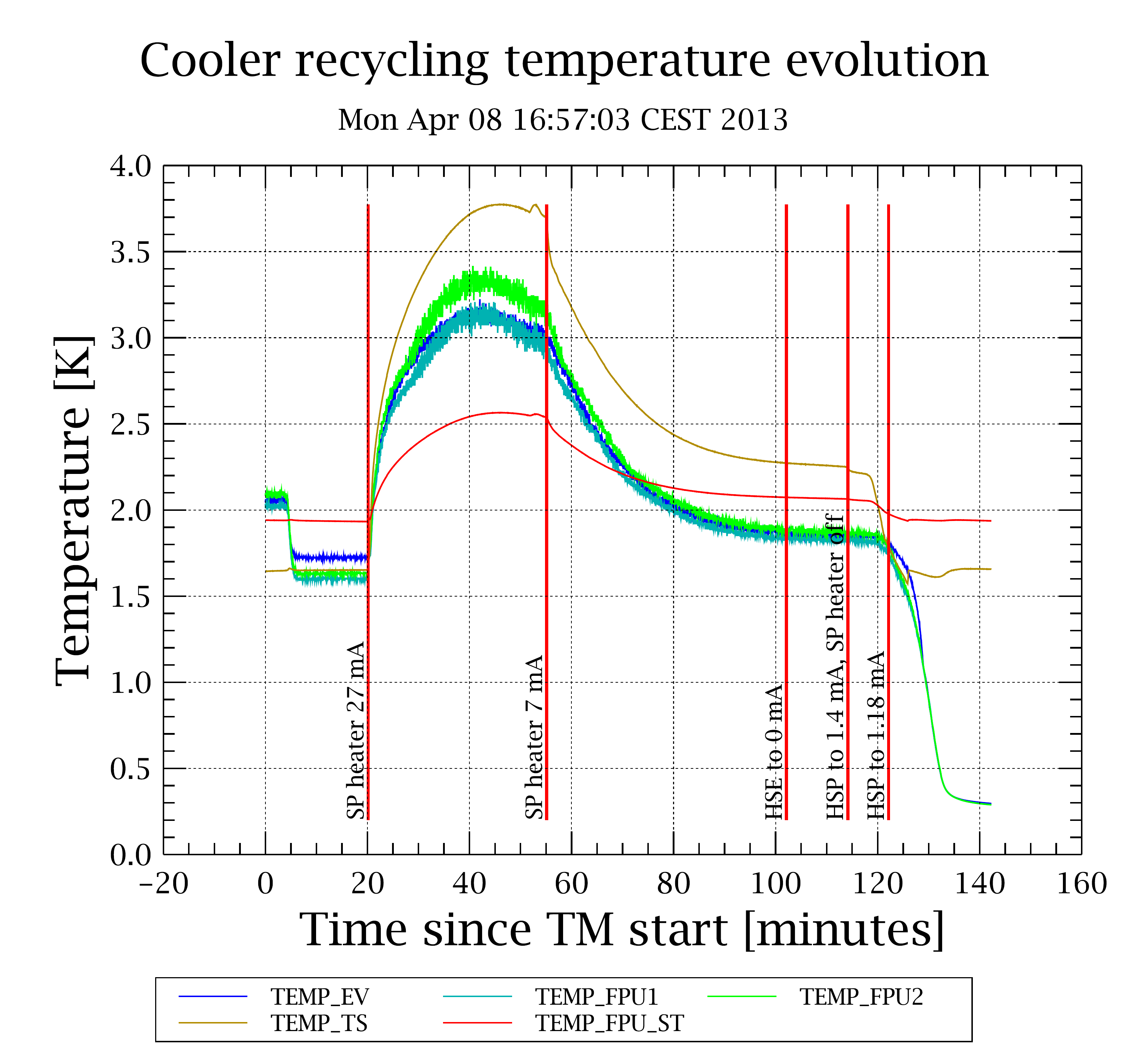}
  \end{center}
\caption{Evolution of temperatures relevant for the PACS cooler
monitored by sensors and provided in the PACS instrument
Housekeeping (HK) during the cooler recycling process. 
}
\label{fig:recycling_temperatures}       
\end{figure}

Fig.~\ref{fig:recycling_temperatures} shows the evolution
of the temperatures relevant for the cooler during the different 
steps of the recycling process as monitored via Housekeeping (HK)
parameters from temperature sensors. At the beginning,
the evaporator temperature (TEMP\_EV) is around 2\,K
indicating that the cooler has run out of liquid coolant. The recycling 
procedure is completely controlled via heaters and heat switches 
integrated in the cooler. The heat switches are of the gas-gap type, 
where the presence or absence of gas between two interlocked 
copper parts changes drastically the heat flow between them. Gas 
handling is achieved with a miniature cryogenic adsorption pump 
(\cite{RefDuband95}).
  
The first 15~minutes serve to settle the thermal environment.
The pump heat switch (HSP in Fig.~\ref{fig:cooler_schematic})
has to be open, so that the pump does not dissipate heat
into the instrument. The evaporator heat switch (HSE 
in Fig.~\ref{fig:cooler_schematic}) is closed so that the
evaporator thermalizes with the 2\,K level and does not warm
up too much when the pump is heated and thus still traps
condensing gas. The heat switches HSP and HSE are closed 
by applying a current and opened by setting the current to 
zero.

The recycling proper starts with heating the sorption pump (SP)
to desorb the gas that has been trapped in the active charcoal.
After 35\,min the heater current is lowered to keep the pump at the
required temperature of about 40\,K. At this time $^3$He
out-gases from the pump and condenses. TEMP\_EV rises
following the temperature rise of the pump both because the enthalpy
of the hot gas coming from the pump is not fully  removed
by the thermal shunt and due to the latent heat of $^3$He.
Since the evaporator and the shunt are connected to the same 
thermal strap, all variations happening at the shunt are also
registered by the evaporator temperature sensor (TEMP\_EV).
TEMP\_EV decreases when the pump is kept at the required 
temperature. At around 80\,min TEMP\_EV drops below the 2\,K 
level. The finally achieved TEMP\_EV in this step characterizes 
the efficiency of the recycling.

After 82\,min the evaporator heat switch (HSE 
in Fig.~\ref{fig:cooler_schematic}) is opened (HSE = 0\,mA) to 
thermally isolate the evaporator. To establish the cooling
functionality the pump heater is switched off after 94 min
and the pump is re-connected to the 2\,K level by closing the pump
switch (HSP in Fig.~\ref{fig:cooler_schematic}).
Once the pump is connected to the 2\,K level the charcoal
in the pump starts pumping and the $^3$He pressure drops.
The thermally insulated liquid $^3$He decreases in temperature
to below the 300\,mK level which is reached after around 142\,min.

The cooler recycling procedure with the tuned time steps
as described above proceeded fully automatically. The duration 
of the PACS-only cooler recycling block was 142.37\,min, the
duration of the parallel mode cooler recycling block 172.53\,min. 
This automatic cooler recycling procedure implemented by means
of the {\it Herschel} Common Uplink System (CUS) is documented
in the online appendix section~\ref{sec:cooler_cus}.

\section{Establishment of the Cooler Hold Time Relation for {\it Herschel}
  Mission Planning}
\label{sec:holdtime_relation}

The {\it Herschel} Space Observatory executed its observations
autonomously along a Mission Timeline (MTL) stored on-board, while
ground contact was only during the normally 3\,h long Daily
Telecommunication Period (DTCP). The mission was divided into Operational
Days (ODs) of on average approximately 24\,h duration, each beginning
with a DTCP during which the MTL for a forth-coming OD was uploaded.
This pre-planned automatic command execution without the possibility
of human intervention meant that a reliable prediction of the
cooler hold time for PACS photometer operations was necessary.
The cooler hold time was defined as the period between TEMP\_EV
going below 300\,mK at the end of the recycling process 
(cf.\ Sect.~\ref{sec:cooler_recycling} and
Fig.~\ref{fig:recycling_temperatures}) and TEMP\_EV exceeding 320\,mK 
when liquid Helium in the cooler was exhausted.

The essential parameter determining the length of the cooler hold
time was the time when the PACS bolometer detectors were biased
for measurement. This period was set by the so-called orbit
prologue and orbit epilogue, engineering-type Astronomical
Observations Requests (AORs) setting and resetting the detector
bias and bracketing the sequence of science AORs. The so-called
biased time was defined as the period between the start of an
orbit prologue and the end of the subsequent orbit epilogue. Note, 
that several biased periods were possible during one single cooler hold time
and the biased time quoted in Table~\ref{tab:cooler_statistics_1}
is the sum of all periods. From OD\,128 onwards, the final optimum 
bias voltages of the PACS detectors were always applied, but on 
earlier ODs of the Performance Verification and Commissioning Phases 
bias settings were varied during detector characterization and optimization
periods, which can have some impact on the resulting hold time. 

\begin{figure}[h!]
  \begin{center}
  \includegraphics[width=1.0\textwidth]{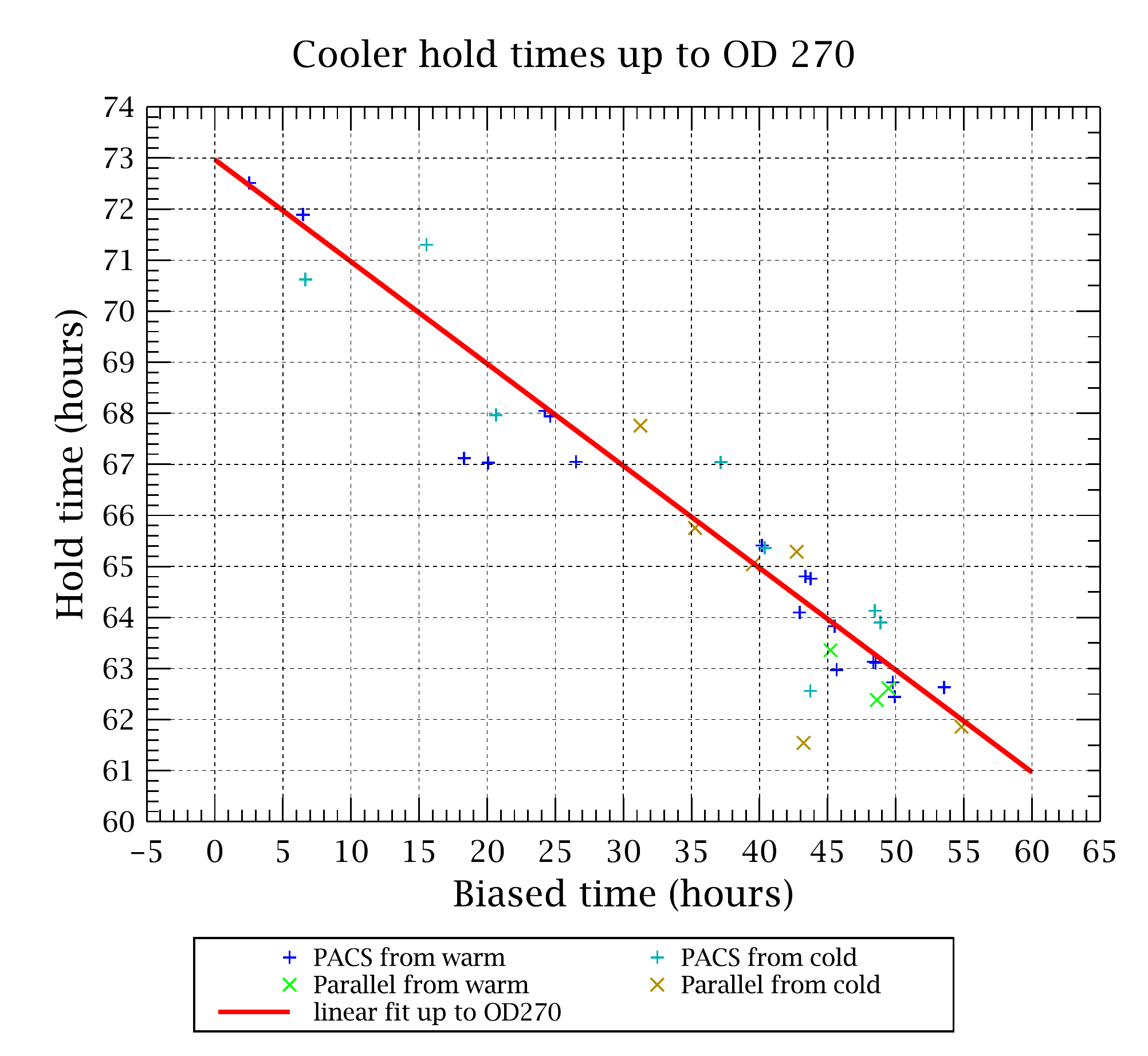}
  \end{center}
\caption{Relation of the PACS cooler hold time with the operational
time of the PACS bolometer detectors for all cooler periods up to
OD\,270 shown as the red line (cf.\ Eqn.(1)). Hold time and biased
time are defined in the text at the beginning of 
Sect.~\ref{sec:holdtime_relation}. Different
symbols and colors represent PACS only or parallel mode cooler
recyclings and the start conditions from a warm, i.e.\ exhausted
liquid $^3$He, or a cold, i.e. still available liquid $^3$He,
cooler.
}
\label{fig:holdtime_relation}       
\end{figure}

Such a dependence was monitored right from the beginning of the
mission and after about three quarters of a year there was
enough statistics to derive a relation as shown in 
Fig.~\ref{fig:holdtime_relation}. The relation \\
\begin{equation}
t_{\rm hold} (h) = 72.97\,h - 0.20 \times t_{\rm bias} (h)
\end{equation} \\

\noindent
established from all complete cooler periods up to OD\,270,
irrespective of the start conditions or whether it was
a PACS only or parallel mode cooler recycling,
was reliably used in the scientific mission scheduling
until the end of the mission with a last PACS cooler
recycling on OD\,1443. The formula was applied for
determination of the cooler hold time following both 
a PACS only and a parallel mode (together with the SPIRE 
cooler) recycling.

In eqn.(1) t$_{\rm hold}$ can be written as 
\begin{math}
t_{\rm hold} = t_{\rm idle} + t_{\rm bias}
\end{math} 
with t$_{\rm idle}$ being the fraction of the hold time
period with the PACS bolometer detectors not biased,
e.g.\ during spacecraft operational maintenance windows
or SPIRE only operations in parallel mode.

In case t$_{\rm idle} \approx$0, then eqn.(1) can be
re-written as \\
\begin{equation}
t_{\rm bias,max} (h) = \frac{72.97}{1.2} (h) = 60.8\,h
\end{equation} \\

\noindent
This meant that a contiguous block of about 
2.5\,ODs of photometer observations could be 
scheduled, thus minimizing the cooler recycling
frequency. The remaining 0.5\,OD was usually 
filled with PACS spectrometer observations, not 
requiring $^3$He cooling, by switching between
the two PACS sub-instruments.

\begin{figure}[hb!]
  \begin{center}
  \includegraphics[width=0.49\textwidth]{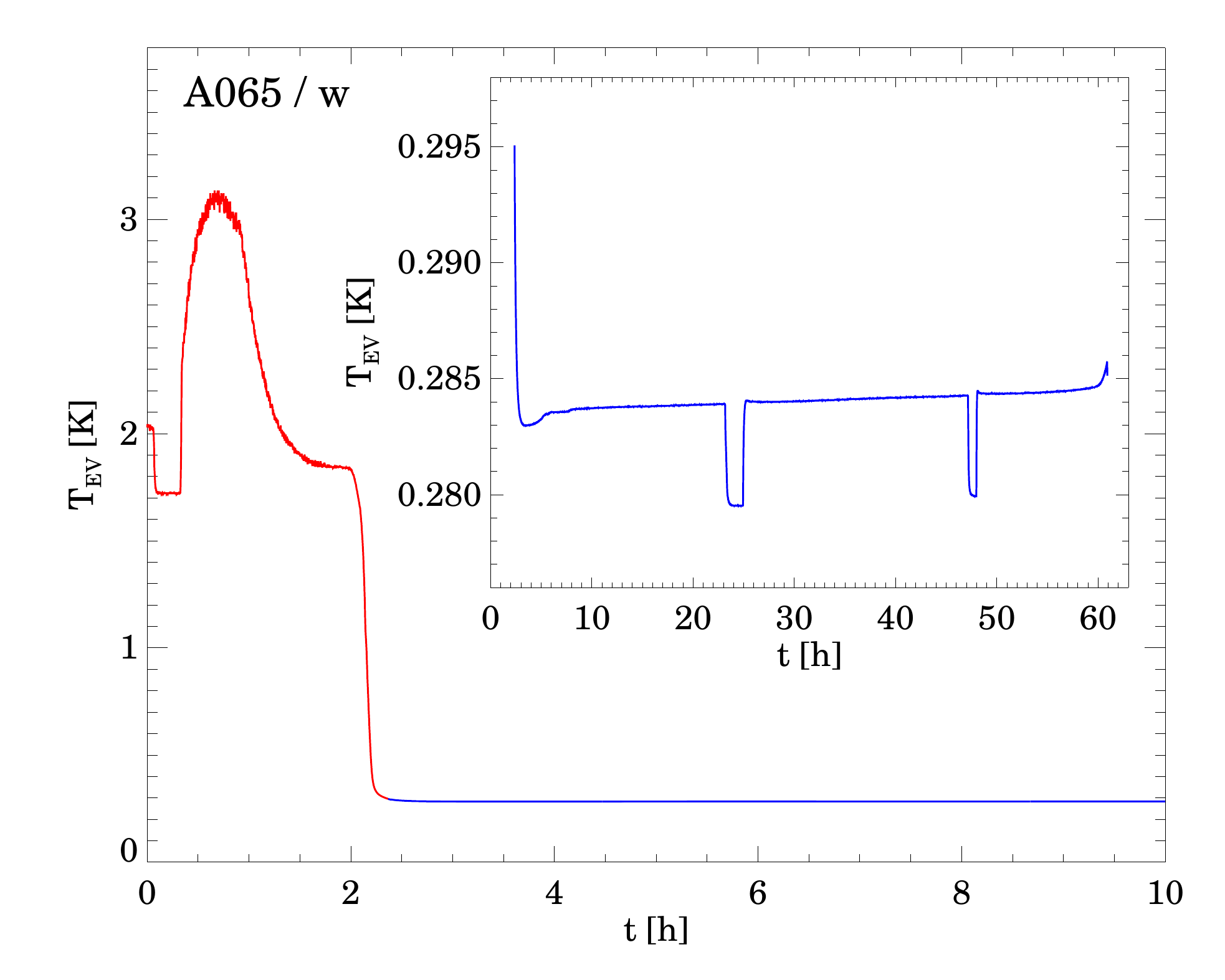}
  \includegraphics[width=0.49\textwidth]{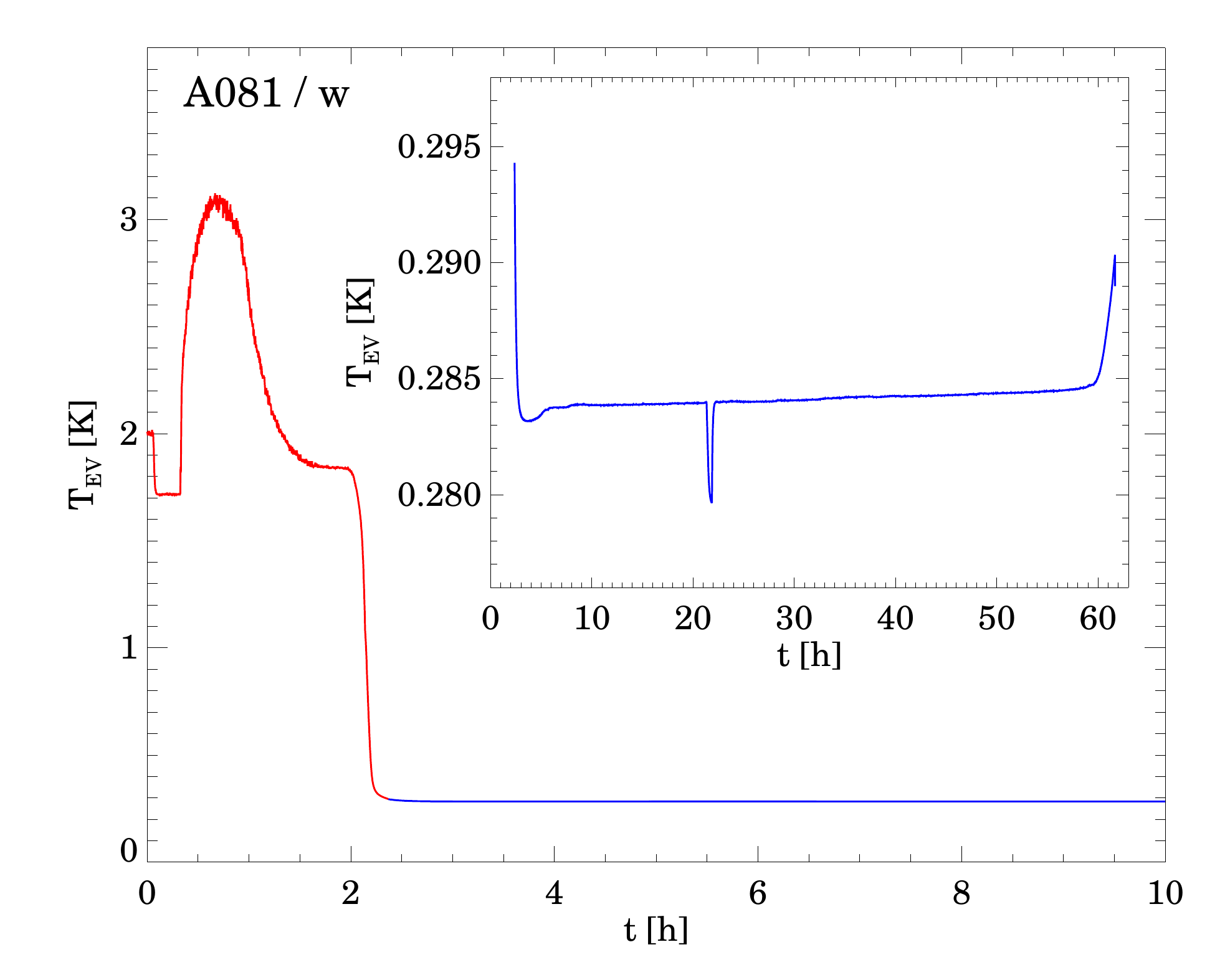}
  \includegraphics[width=0.49\textwidth]{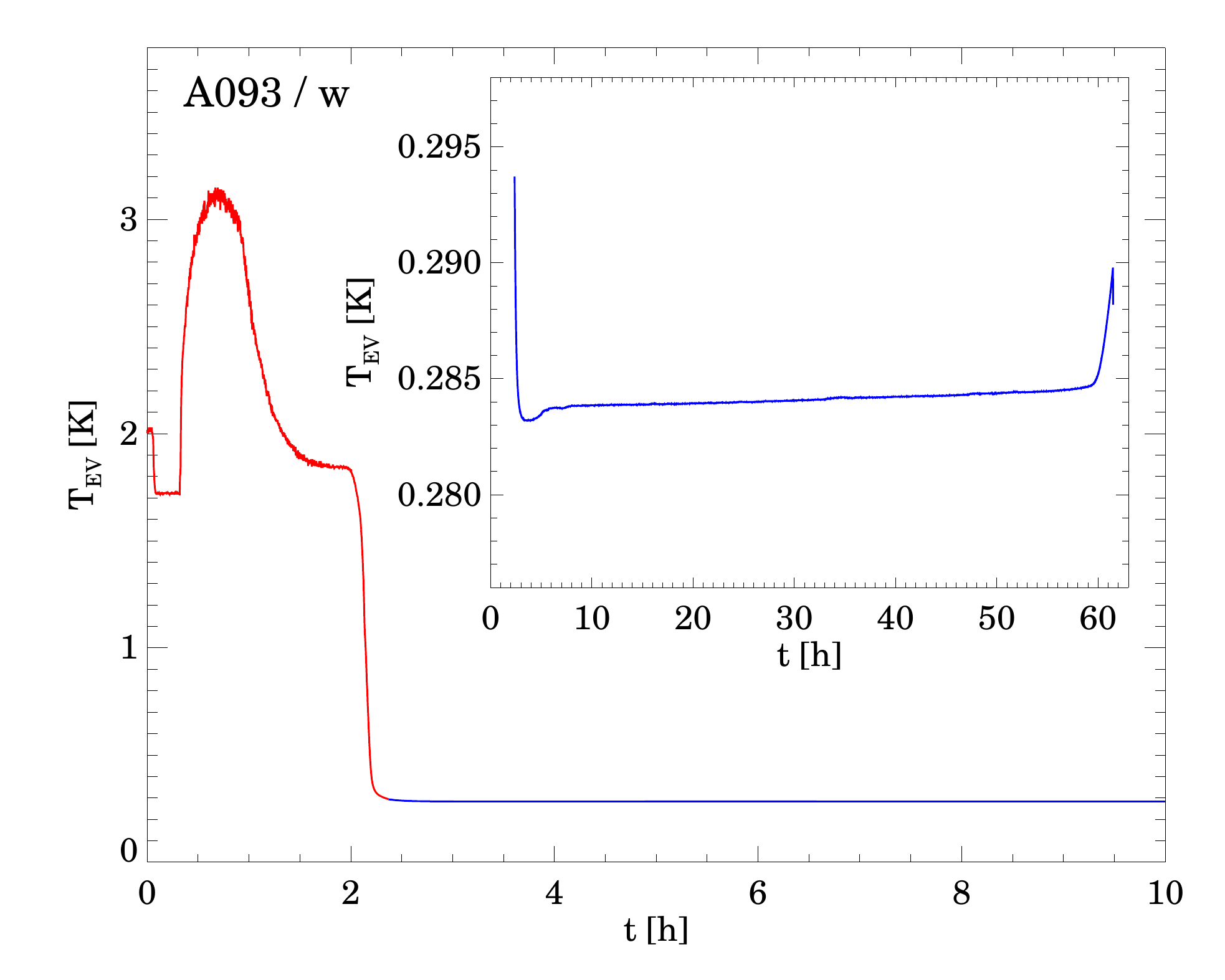}
  \includegraphics[width=0.49\textwidth]{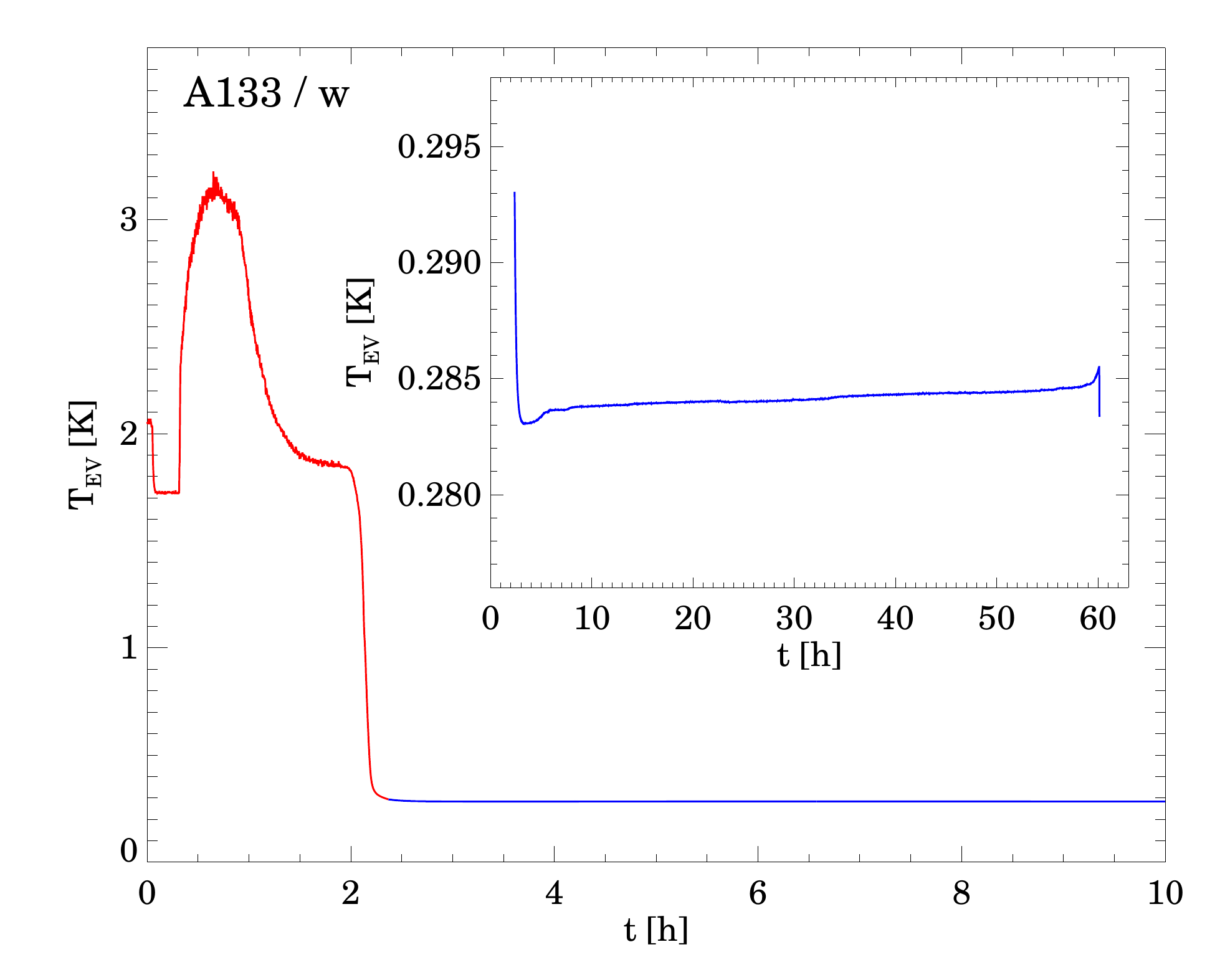}
  \end{center}
\caption{Individual PACS-only cooler cycles (labeled ``A''
plus a sequence number, see Table~\ref{tab:cooler_statistics_1})
representing the four different operational periods with 
regard to biasing the detectors. The figure in the main 
panel shows the course of the evaporator temperature
(T\_EV) during the first 10\,h following the start of 
the recycling, whereby the red part represents the 
proper recycling process, the blue part the beginning 
of the subsequent operational period. The inserts are 
a zoomed view with adapted dynamic range of T\_EV over 
the full operational period. Top left: Cycle A065 on 
OD\,469 with 3 bias periods, the drops of T\_EV to the 
280\,mK level indicate the unbiased state of the detectors;
top right: Cycle A081 on OD\,684 with 2 bias periods;
bottom left: Cycle A093 on OD\,842 with one maximum 
contiguous bias period of nearly 59.2\,h; bottom right:
Cycle A133 on OD\,1354 with one reduced contiguous
bias period of 57.8\,h. For the latter one the final
steep temperature increase is less than 1\,mK. 
}
\label{fig:cooler_cycles}       
\end{figure}

Since the cooler hold time relation in eqn.(1)
was determined for the point in time when the
evaporator temperature exceeded an out-of-limit
value of 320\,mK, a safety time buffer was included
when using this relation for mission planning,
thus reducing t$_{\rm bias,max}$ by t$_{\rm buffer}$.
Since the evaporator temperature increased
very steeply only at the end of the cooler hold time
(see Fig.~\ref{fig:cooler_cycles}),
a buffer of 1.5\,h was deemed sufficient initially.
However, inspection of photometer calibration
observations, scheduled deliberately at the end of 
the cooler period for cross-calibration with 
subsequently scheduled PACS spectrometer observations
of the same target, showed that there was already
an increase in the evaporator temperature (cf.\ 
the course of the temperature in Fig.~\ref{fig:cooler_cycles}
with an increase of up to 5\,mK for the two cycles
labeled A081 and A093; for the naming convention see beginning
of Sect.~\ref{sec:statistics} and online Table~\ref{tab:cooler_statistics_1})
with an impact on the photometric calibration accuracy 
(\cite{Refbalog13},~\cite{Refnielbock13}). 
This effect is described in detail and characterized
by appropriate correction functions in~\cite{Refmoor13}.
Despite the final calibration and correction of this
effect, it was decided to increase t$_{\rm buffer}$
to 3\,h, which meant that the evaporator temperature
increased by at most 1\,mK at the end of the 
cooler period. This meant a maximum contiguous
PACS photometer operation of 57.8\,h.

\section{Mission Statistics of Cooler Recycling}
\label{sec:statistics}

A total number of 239 PACS cooler recyclings was
performed during the {\it Herschel} mission,
the first one on OD\,26 for PACS commissioning and
the last one on OD\,1443. 139 cooler recyclings
were performed as PACS only (labeled ``A'' plus a 
sequence number), 100 recyclings in parallel
with the SPIRE cooler recycling (labeled ``B'' plus
a sequence number). It should be noted
that the parallel cooler recyclings were not only performed
for parallel mode observations of PACS and SPIRE, but also 
for separate PACS and SPIRE observations during the 
subsequent hold time periods. The SPIRE cooler hold time
was restricted to about 48\,h, the difference between
the PACS and SPIRE cooler hold times was usually covered
by PACS photometer observations to achieve the highest
possible efficiency of each cooler recycling
with regard to the foreseen and available photometer 
observing program.

Online Table~\ref{tab:cooler_statistics_1} gives an overview of
all cooler cycles with the achieved hold time
and the duration of the PACS detector operations
during this hold time. It can be recognized that
in the beginning of the mission the detector operational
times were relatively short owing to the step-by-step 
commissioning and performance assessment of the PACS 
photometer. From about OD\,200 onwards, routine observations
started achieving usually 40 -- 50\,h of photometer 
operations, but this depended also on the instrument
allocation per Operational Day due to target visibility
and priority of the observations. In the beginning,
the PACS photometer was operated in a safe manner,
it was switched into its standby mode by an orbit 
epilogue AOR, i.e. detectors were de-biased, at the end 
of each OD and whenever another instrument became prime. 
This is reflected in the usual 2 -- 3 bias periods during 
this time. This safety aspect could be relaxed along the 
mission due to the smooth operation of the {\it Herschel} 
Space Observatory. From OD\,675 onwards, the PACS photometer 
was switched to standby mode after the first OD of a photometer 
block, but was then left in biased mode for the subsequent 
1.5 ODs. From OD\,732 onwards, the PACS photometer was kept 
in biased mode for the full 2.5\,ODs, thus reaching
maximum biased times of up to 59.2\,h, and later 
57.8\,h after increasing the buffer time at the end of the 
hold time (see Sect.~\ref{sec:holdtime_relation}).
From OD\,824 onwards, PACS and SPIRE were operated 
in most cases simultaneously in prime mode following a 
parallel cooler recycling without interference in their HK 
telemetry. This saved time, which had to be spent for the 
instrument set-up from standby into biased mode, for science 
observations, but did not mean full permanent observations 
with PACS during the biased time due to spacecraft operations 
windows or SPIRE only observations. Nevertheless, this 
final way of operation yielded the most optimum usage of 
the PACS and SPIRE photometers for science observations. 

Fig.~\ref{fig:cooler_cycles} shows individual cooler cycles
from these different modes of operating the detectors.
The periods of unbiased detectors can be clearly recognized
by a temperature drop of about 4\,mK. After biasing the 
detectors again, the temperature goes up to the previous 
level. There is a relatively smooth slight increase
of the evaporator temperature during the cycle until the
very end when the temperature rises much more steeply. 
Depending on the safety buffer, this final increase ranges
from less than 1\,mK for the 3\,h buffer to about 5\,mK
for the 1.5\,h buffer. 

\begin{figure}[ht!]
  \begin{center}
  \includegraphics[width=0.49\textwidth]{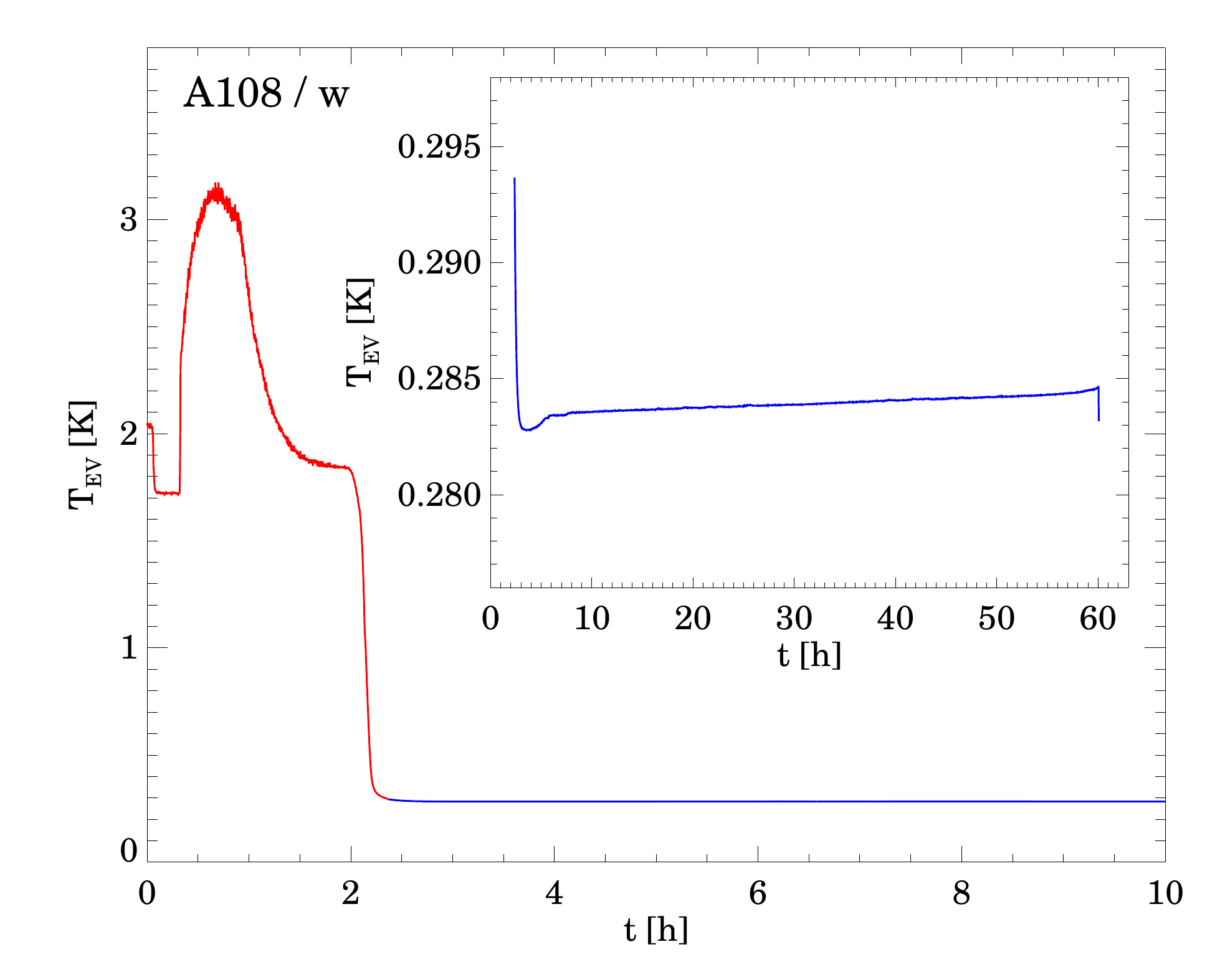}
  \includegraphics[width=0.49\textwidth]{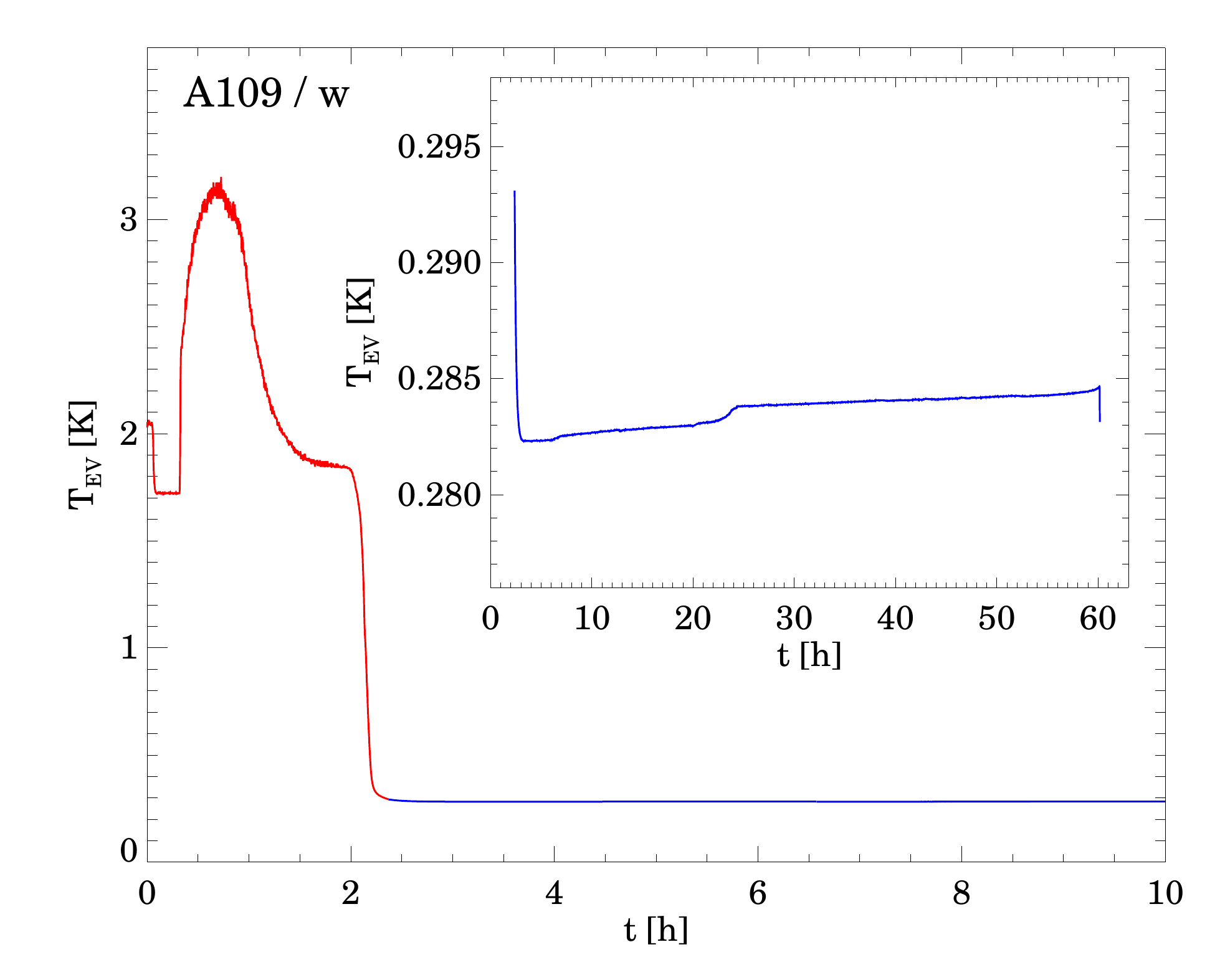}
  \end{center}
\caption{Examples of the two main types of evaporator
temperature evolution during a cooler cycle not disturbed
by switching detector biases. Left: Initial dip followed
by a flat slope. Right: ``Swell'' (shoulder with
a temperature rise of about 1\,mK around 24\,h after
the start of the recycling). The lay-out and labeling of 
the panels is the same as in Fig.~\ref{fig:cooler_cycles}.   
}
\label{fig:cooler_bump}       
\end{figure}

When inspecting the shape of the cooler evaporator temperature
along the cooler cycle, we can distinguish two main types,
as illustrated for two subsequent cycles undisturbed by detector 
bias switching in Fig.~\ref{fig:cooler_bump} (this
characteristic difference applies in general also for cycles with 
detector bias switching).

\begin{itemize}
\item[1)] Following the drop to the operational temperature
          between 282 and 283\,mK (if detectors are already
          biased), the evaporator temperature shows a flat
          increase. Around 24\,h after the start of the
          cooler recycling, there is a 1\,mK high swell
          in the temperature curve.
          This usually occurs at around 24\,h (not always at
          exactly the same time), but there are some cases when 
          it already occurs after less than 10\,h or only after 
          more than 30\,h.
\item[2)] Directly after the drop, there is an initial dip
          followed by the flat temperature rise. A swell 
          does not occur during this cycle.
\end{itemize}

Most of the cooler cycles, both PACS only and parallel,
show the ``swell'' shape. Column 9 in Table~\ref{tab:cooler_statistics_1}
indicates the shape for each cooler cycle. ``Swell''-type
cycles start at a lower temperature level than ``dip''-type
cycles and reach the temperature level of the ``dip''-type cycle
at the time of the ``swell''. The duration of the ``swell''
is 2-3\,h. The ``dip'' has a similar duration.

\begin{figure}[h!]
  \begin{center}
  \includegraphics[width=0.86\textwidth]{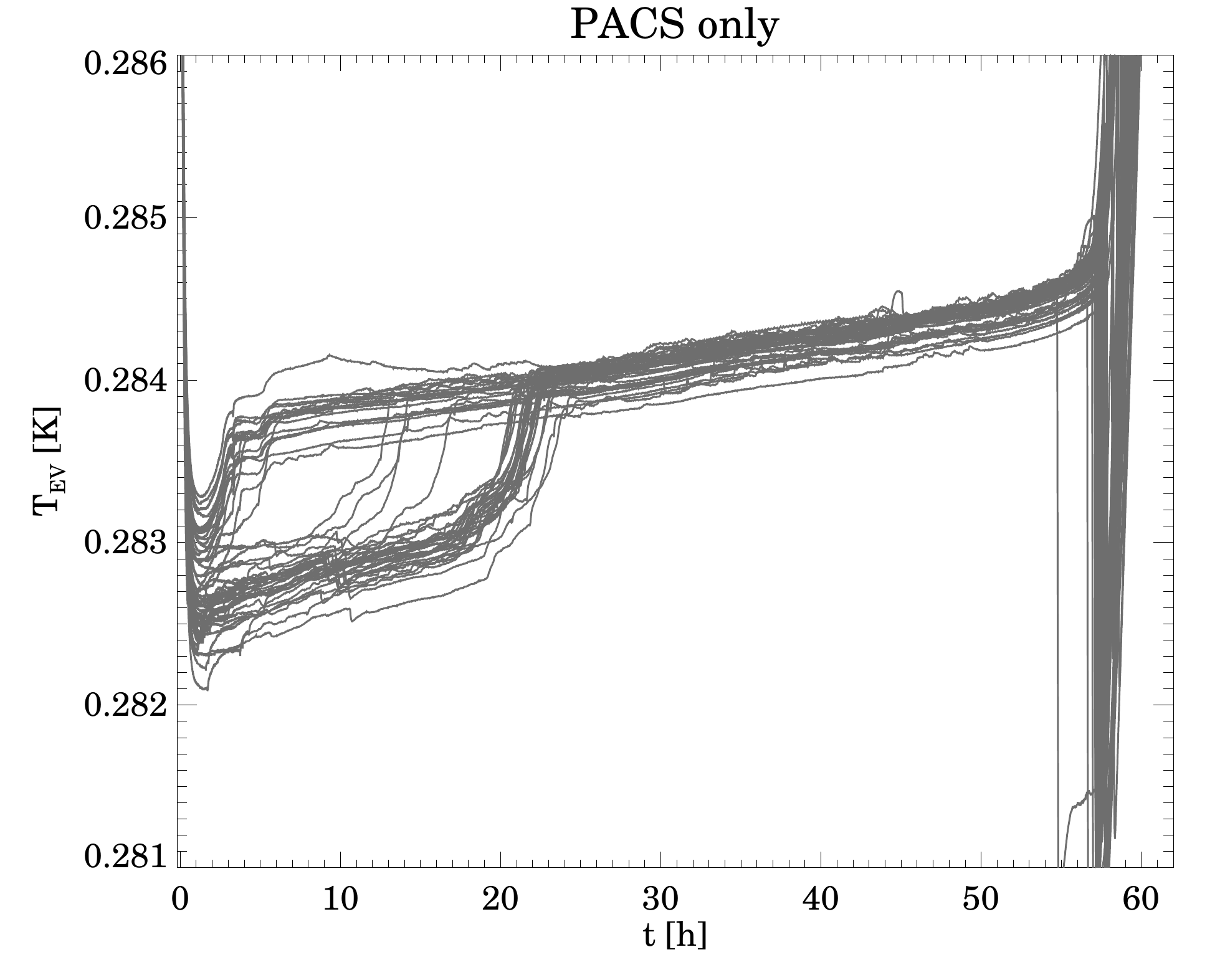}
  \includegraphics[width=0.86\textwidth]{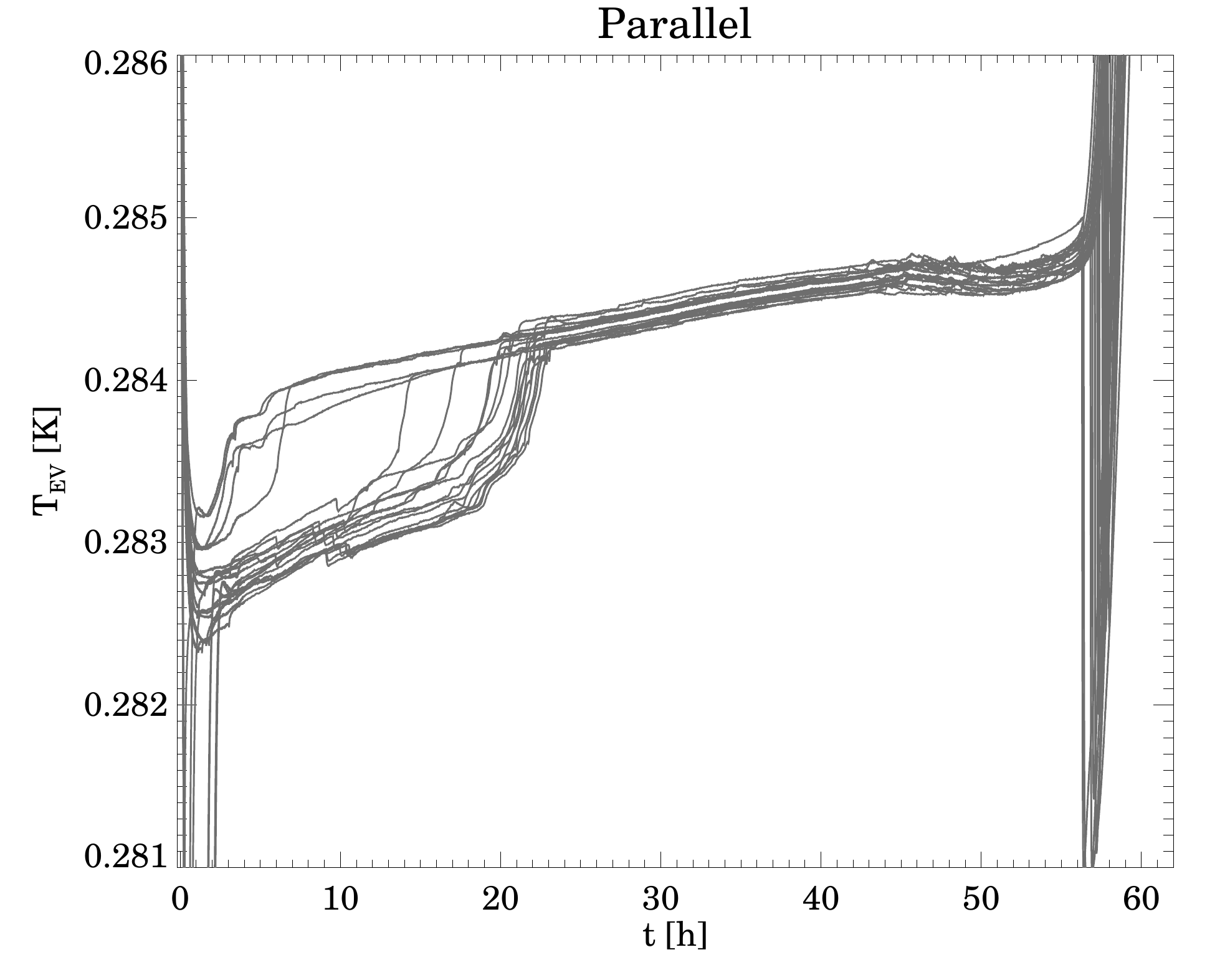}
  \end{center}
\caption{Superposition of all cooler cycle evaporator
temperature curves with full cooler hold time outlining 
the main characteristics and variation of the two main 
evaporator temperature evolution families as described 
in the text. Top: PACS-only cycles. Bottom: Parallel cycles. 
The zero point in time is the end of the preceding cooler 
recycling. 
}
\label{fig:cooler_curves}       
\end{figure}

As a summary of the described characteristics of the
evaporator temperature evolution, Fig.~\ref{fig:cooler_curves}
shows a superposition of all cooler cycles with full hold
time and without de-biasing in-between, separately for
PACS-only and parallel mode cycles. This reveals fine 
differences for the two modes of operations. Parallel
cycles show an on average slightly higher temperature level, 
which indicates some small impact by the parallel operations 
of the SPIRE instrument. This impact is confirmed by the temperature
drop after about 48\,h, when SPIRE operations terminate. Thanks 
to the photometric calibration method correcting for time 
dependent evaporator temperature variation (\cite{Refmoor13}), 
all photometric measurements are put on a homogeneous 
temperature reference level.

The ``dip-type'' evaporator temperature curves
look similar to the observed on-ground behavior.
The ``dip'' appears therefore to be a consequence of 
the relative timing of the end of the cooler recycling
procedure, when the evaporator temperature still
drops, and the bias setting of the detectors
as part of the orbit prologue procedure, which
leads to the turn of the temperature curve to the
level for biased detectors.

We have investigated whether the evaporator temperature
``swell'' is related to instrument or satellite specific
activities. We could not find any strong correlation
with instrument switching, long slews, in particular the
concluding ones of an OD, or spacecraft operational windows
for reaction wheel de-biasing. The physical reason
behind the occurrence of the ``swells'' in the evaporator
temperature remains therefore an open issue. We believe it 
to be a zero gravity effect, possibly related to the liquid
arrangement inside the porous material, since it was never
observed during tests of the cooler system on ground.
Note that for the SPIRE cooler a similar phenomenon has
been reported, which has been called the ``SPIRE cooler burp''.

\begin{figure}[hb!]
  \begin{center}
  \includegraphics[width=0.49\textwidth]{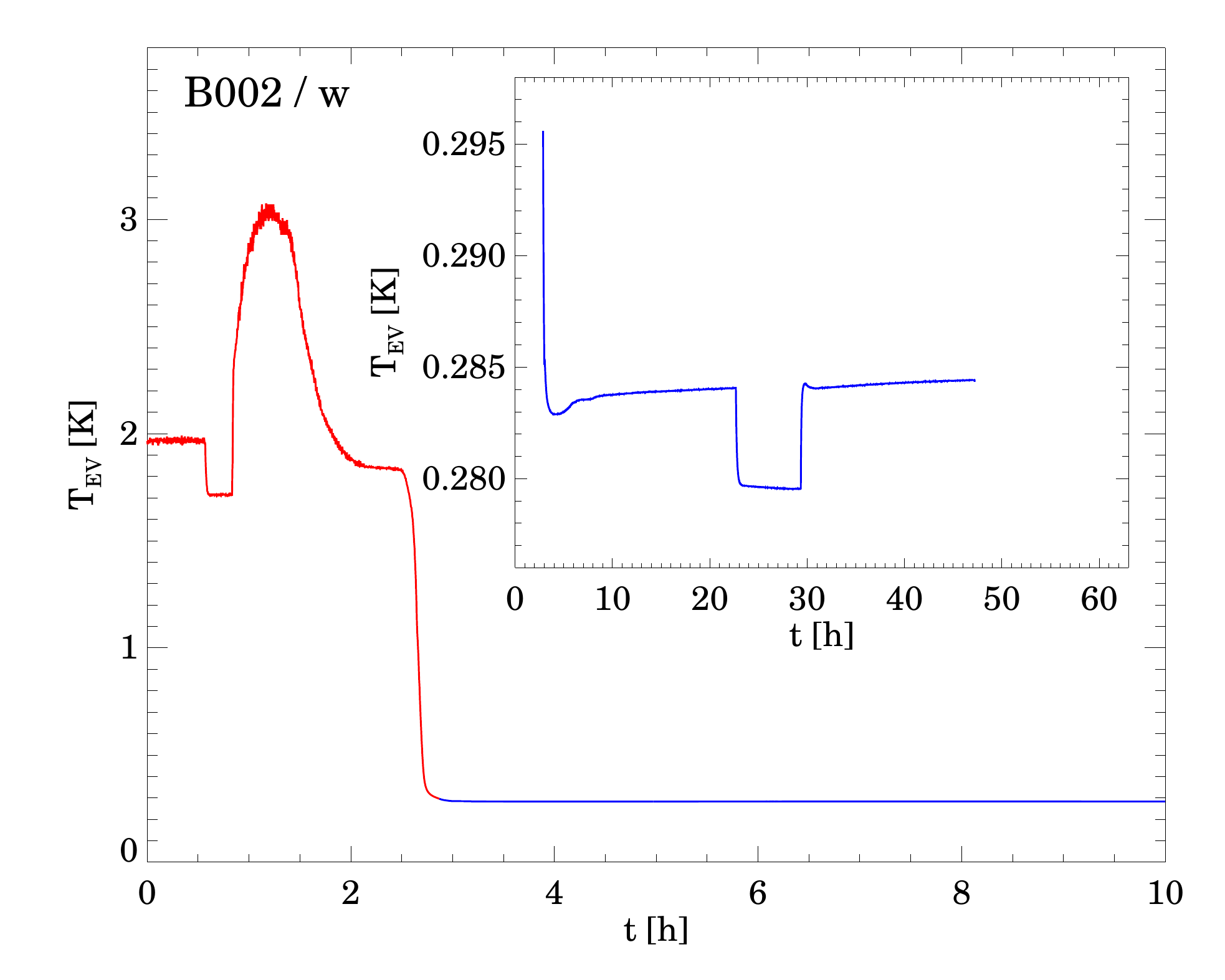}
  \includegraphics[width=0.49\textwidth]{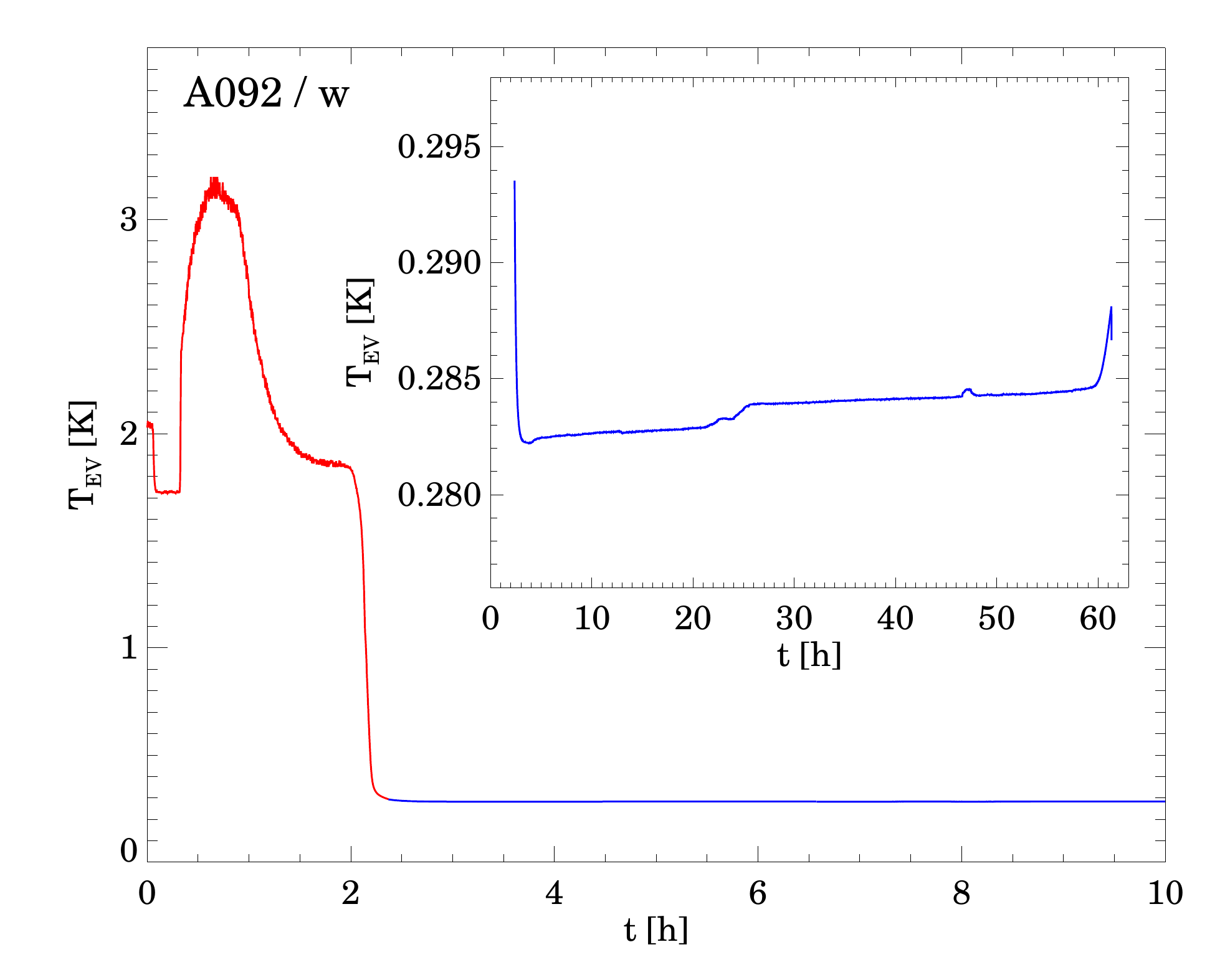}
  \end{center}
\caption{Examples of cooler cycles with particular detector
operations. Left: During OD\,137 (2nd half of the cycle
after the temperature rise terminating the unbiased period,
between 38 and 47\,h) Mars was illuminating the detector 
pixels for 9\,h. Right: During ODs 831 and 832 detectors 
were operated in a specific engineering mode using a different
detector bias setting. These periods are visible as two humps, 
one on top of the evaporator temperature ``swell'', the other 
at around 47\,h. The lay-out of the panels is the same as in 
Fig.~\ref{fig:cooler_cycles}. Labels ``A'' and ``B'' with a
sequence number represent a PACS-only or a parallel mode cycle,
respectively.
}
\label{fig:detector_operations}       
\end{figure}

We also investigated the impact of the detector operations
on the shape of the evaporator cooler temperature curve.
This is illustrated in Fig.~\ref{fig:detector_operations}.
Even with the brightest measurable sources on the detector,
the illumination did not cause any noticeable deviations
of the evaporator temperature. A noticeable effect could be
recognized when the detectors were operated with different
bias settings, but this was restricted to a few engineering 
and calibration observations only. 

\begin{figure}[h!]
  \begin{center}
  \includegraphics[width=1.0\textwidth]{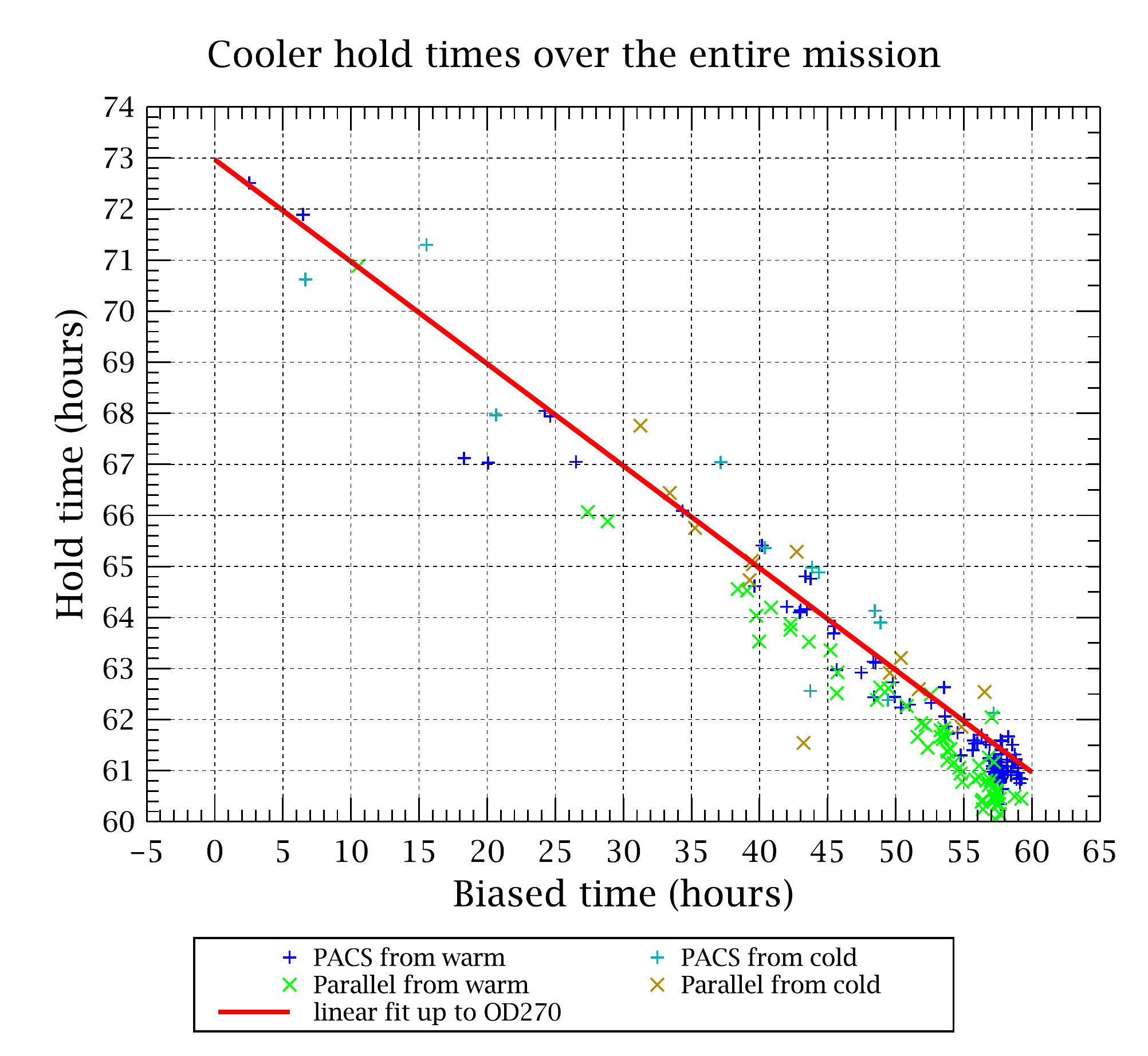}
  \end{center}
\caption{Statistics of the cooler hold time versus bolometer
biased time over the entire {\it Herschel} mission. Hold time 
and biased time are defined in the text at the beginning of 
Sect.~\ref{sec:holdtime_relation}. Different
symbols and colors represent PACS only or parallel mode cooler
recyclings and the start conditions from a warm, i.e.\ exhausted
liquid $^3$He, or a cold, i.e. still available liquid $^3$He,
cooler. The operational guideline established from the relation 
of all cooler periods up to OD\,270 (cf.\ Fig.~\ref{fig:holdtime_relation})
is shown as the red line. 
}
\label{fig:holdtime_mission}       
\end{figure}

Fig.~\ref{fig:holdtime_mission} shows the relation of the
cooler hold time versus the biased time for all complete
cooler cycles during the mission (192 out of 239, 115 out of
139 PACS only, 77 out of 100 parallel). Cooler periods flagged
in Table~\ref{tab:cooler_statistics_1} by a banner ``$^{t}$'' 
in the column hold time were not considered, because of their 
truncation by the start of a new cooler recycling under 
cold start conditions before the final exhaustion of the 
liquid $^3$He.

Plotting all complete hold times of the entire mission
shows some fine differences between the different modes
and start conditions. The PACS-only recyclings starting
from a warm cooler cluster close to the relation 
established from all complete cooler periods up to OD\,270 (cf.
Sect.~\ref{sec:holdtime_relation} and Fig.~\ref{fig:holdtime_relation},
as represented by the red line in Fig.~\ref{fig:holdtime_mission} 
with a dispersion of $\approx\pm$0.5\,h. The parallel cooler 
recyclings starting from a warm cooler are slightly shifted 
to an about 1\,h shorter cooler hold time with a similar
dispersion. This one hour less efficiency was well covered
by the buffer time applied for mission planning aspects
as described in Sect.~\ref{sec:holdtime_relation}.
Statistics for the cooler recyclings starting
from a still cold cooler are poorer, but there is some indication
that the hold times both for PACS-only and parallel cooler recycling 
are about 0.5\,h longer than for the warm start PACS-only cooler 
recyclings. 

There are a few outliers with hold times shorter
by 1.5 to 3\,h than comparable cycles which are restricted to 
one period between OD\,101 and OD\,132. Note that this
period largely coincides with the period when the bias
voltage settings of the PACS detectors were still being varied for
detector performance optimization (prior to OD\,128). 

Formal linear fits like the one for the general hold time relation in Eqn(1)
yield \\
for~all~cycles:
\begin{equation}
t_{\rm hold} (h) = 72.747\,h - 0.205 \times t_{\rm bias} (h) 
\end{equation}
for PACS-only cycles with warm start: 
\begin{equation}
t_{\rm hold} (h) = 72.523\,h - 0.198 \times t_{\rm bias} (h)  
\end{equation} 
for PACS-only cycles with cold start:
\begin{equation}
t_{\rm hold} (h) = 72.708\,h - 0.187 \times t_{\rm bias} (h) 
\end{equation} 
for parallel cycles with warm start:
\begin{equation}
t_{\rm hold} (h) = 72.351\,h - 0.204 \times t_{\rm bias} (h) 
\end{equation} 
for parallel cycles with cold start:
\begin{equation}
t_{\rm hold} (h) = 72.958\,h - 0.200 \times t_{\rm bias} (h) 
\end{equation} 

\noindent
Investigation of the cause for the dispersion in cooler hold time
for the same biased time has to consider the following aspects:

\begin{itemize}
\item[1)] Short term variations of the thermal environment
         in the order of days.
\item[2)] Systematic trends along the {\it Herschel} mission.
\end{itemize}

\begin{figure}[ht!]
  \begin{center}
  \includegraphics[width=1.00\textwidth]{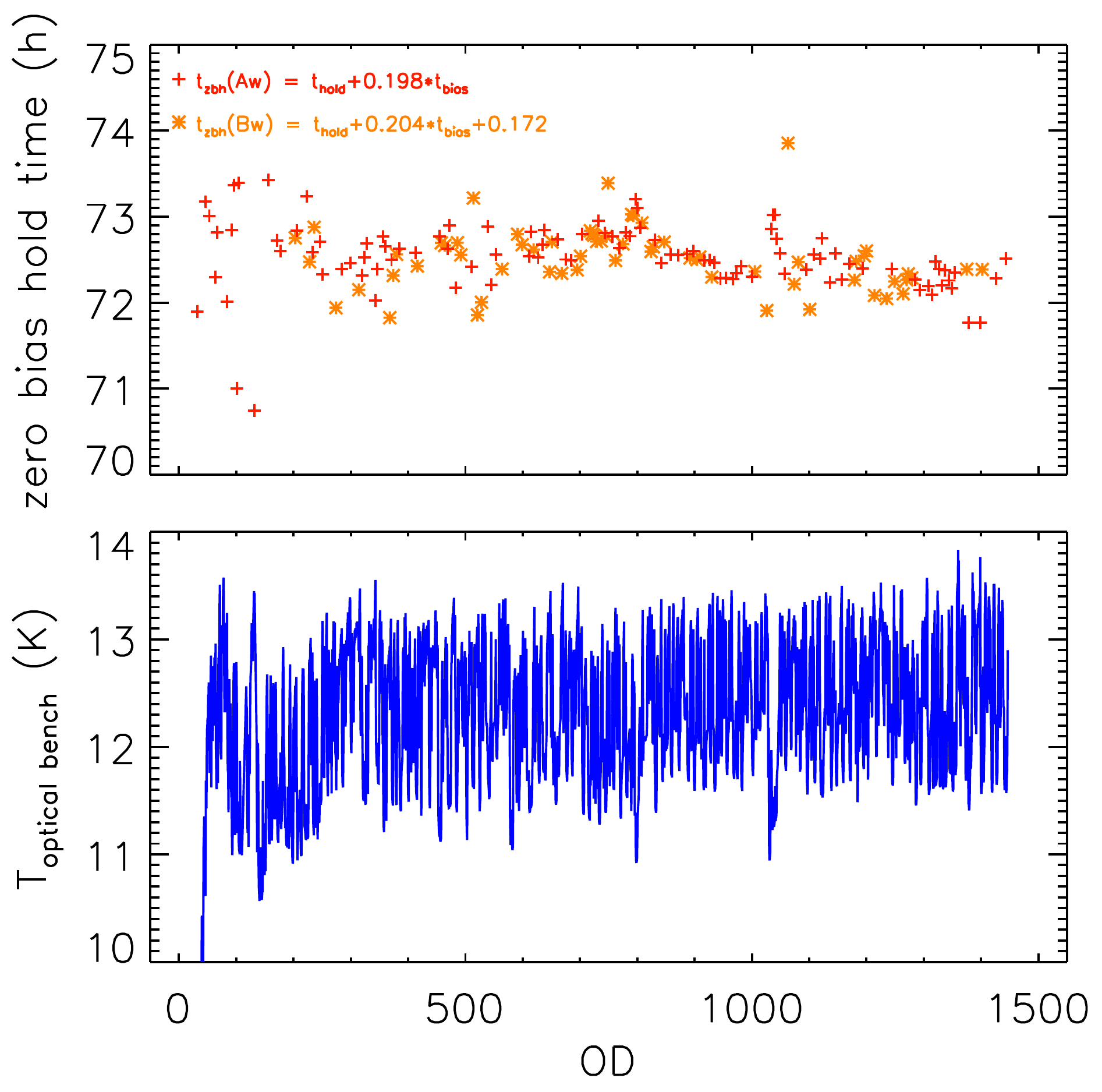}
  \end{center}
\caption{Top: Evolution of the zero bias hold time (definition in the text)
with mission time represented in Operational Days (ODs) for cooler
cycles with a warm (w) start. PACS-only (A) and parallel mode (B) 
cooler cycles are represented by different symbols. The relations
to calculate the zero bias hold time are indicated. Bottom: Temperature 
reading of the PACS optical bench sensor which reflects the total thermal 
input to the PACS focal plane unit.
}
\label{fig:cooler_OD}       
\end{figure}

This investigation is illustrated in Fig.~\ref{fig:cooler_OD}.
For a homogeneous comparison of cooler cycles with very different
biased times we calculate the zero bias hold time $t_{\rm zbh}$ as
\begin{equation}
t_{\rm zbh} (h) = t_{\rm hold} (h) + f \times t_{\rm bias} (h)
\end{equation}

\noindent
with $f$ being the factors in the relations of Eqns. (4) and (6),
respectively. To account for the slightly different average
hold times of PACS only and parallel mode recyclings,
the difference of the offsets (72.523\,h - 72.351\,h = 0.172\,h)
has been added to $t_{\rm zbh}$ of parallel cycles.

\noindent
Fig.~\ref{fig:cooler_OD} features the following characteristics
\begin{itemize}
\item[1)] There is quite some dispersion up to about OD\,200
          including the mentioned outliers.
\item[2)] From OD\,200 onwards there are periods with quite
          low dispersion in $t_{\rm zbh}$ interrupted by periods
          with higher scatter.
\item[3)] There appears to be a slight downward trend of $t_{\rm zbh}$
          along the mission.
\end{itemize}  

In trying to find correlations with instrument temperature sensing,
the PACS instrument optical bench temperature (in the range 10 - 13\,K)
appears to reflect individual features and overall trends 
(cf. Fig.\ref{fig:cooler_OD}). Note that this is an anti-correlation
with low optical bench temperatures corresponding to long hold times
and vice versa. It is not always a 1-to-1 anti-correlation in 
time, but the thermal history also has to be considered.
The PACS optical bench temperature is a representation of the total
thermal load into the PACS instrument. It depends on activities of 
all systems inside the cryostat. Switch-off of one of the
three instruments, e.g. due to contingency mainly by cosmic ray hits, 
can cause a significant drop of the thermal load. The thermal load
and environment can have an impact both on the recycling process
and the evaporation level. We noted some slight systematic variations
of the evaporator temperature curves during the recycling, like
an increase of the peak temperature and the final 2\,K level
of the recycling, with proceeding mission.

\begin{figure}[h!]
  \begin{center}
  \includegraphics[width=1.0\textwidth]{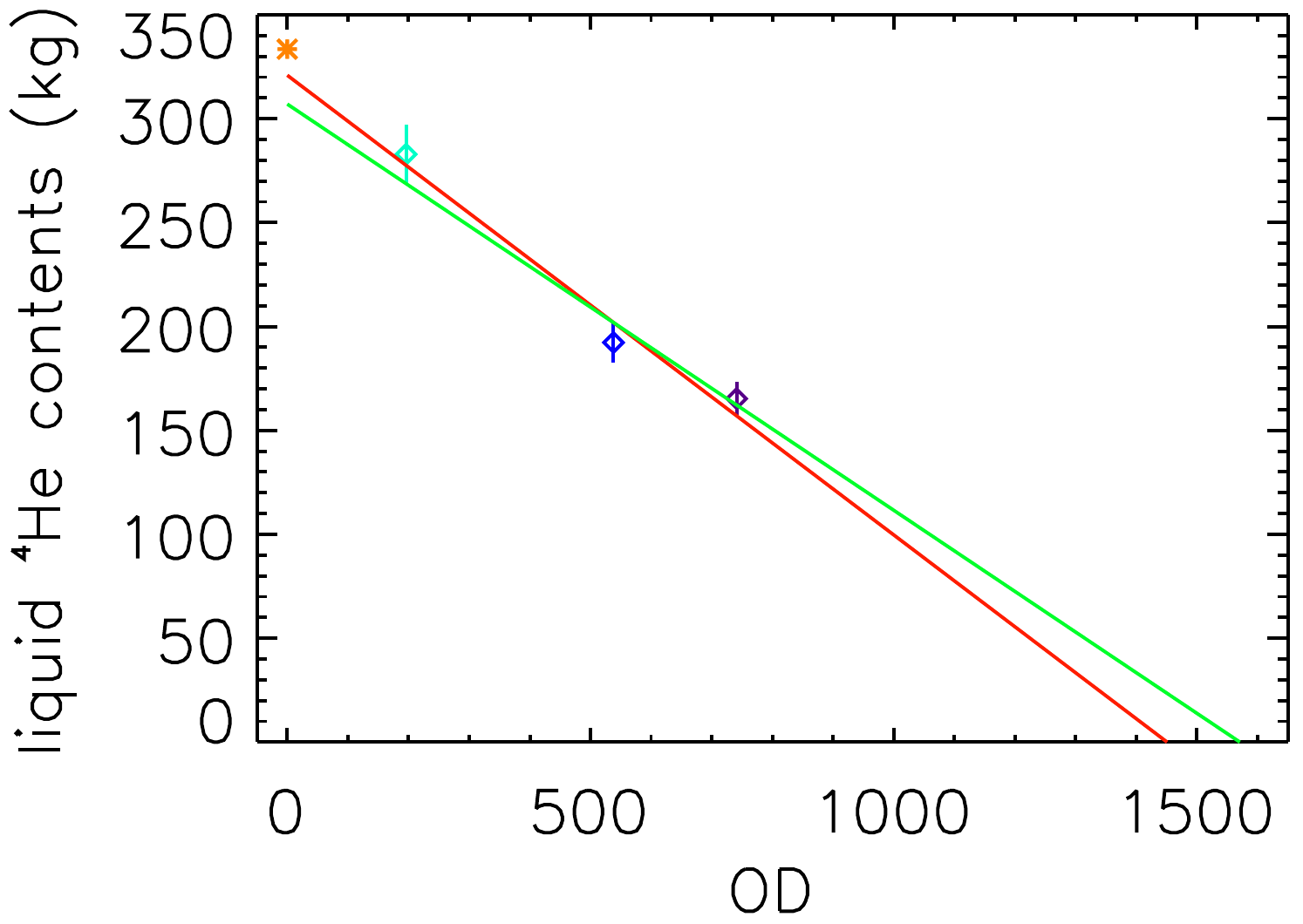}
  \end{center}
\caption{End of Life (EOL) prediction of the superfluid $^4$He content
in the {\it Herschel} cryostat on the basis of three dedicated Direct 
Liquid Content Measurements (DLCM) on ODs\,196, 538, and 741. The 
solid green and red lines give the best and worst linear fit to the
three data points. The point at OD\,0 is the mass estimate after the
final Helium top-up of the Herschel cryostat prior to launch. Due to
the cool-down of the telescope during the first 2 months, the initial 
Helium consumption is different excluding this point from the fit of 
the remaining mass in-orbit (source: ESA). 
}
\label{fig:dlcm-LHelifetime}       
\end{figure}

\section{Utilization of cooler recyclings to determine the lifetime of $^4$He in the {\it Herschel} cryostat}
\label{sect:He_lifetime}

The lifetime of a cryogenic mission is usually determined
by the consumption of the initial mass of coolant. In the
case of {\it Herschel} this was the amount of superfluid
$^4$He carried inside the instrument cryostat. For a realistic
mission planning trying to execute the essential scientific
program a feedback on the remaining lifetime is very helpful.
For that reason the {\it Herschel} cryostat was equipped
with special heaters to perform Direct Liquid Content Measurements 
(DLCM). Three measurements of this type were performed on ODs\,196,
538 and 741 by sending a dedicated heat pulse into the
superfluid $^4$He and watching its thermal reaction.
This meant however each time the interruption of the scientific 
observations for several hours to achieve stable thermal conditions
and the results did not show a smooth gradual reduction of the
$^4$He mass content, leaving a relatively large uncertainty
of about 4 months in lifetime, cf. Fig.~\ref{fig:dlcm-LHelifetime}.

\begin{figure}[h!]
  \begin{center}
  \includegraphics[width=1.0\textwidth]{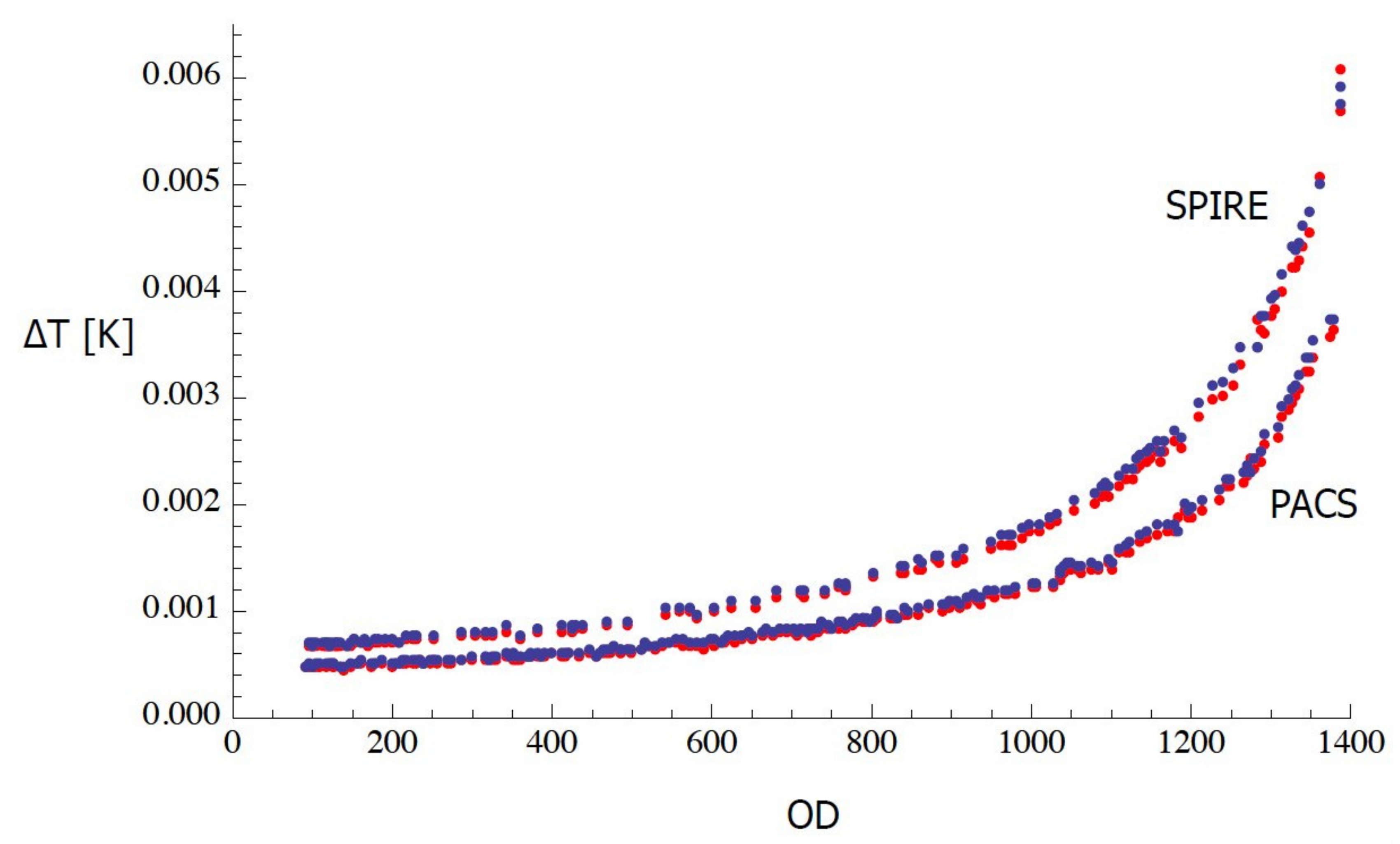}
  \includegraphics[width=1.0\textwidth]{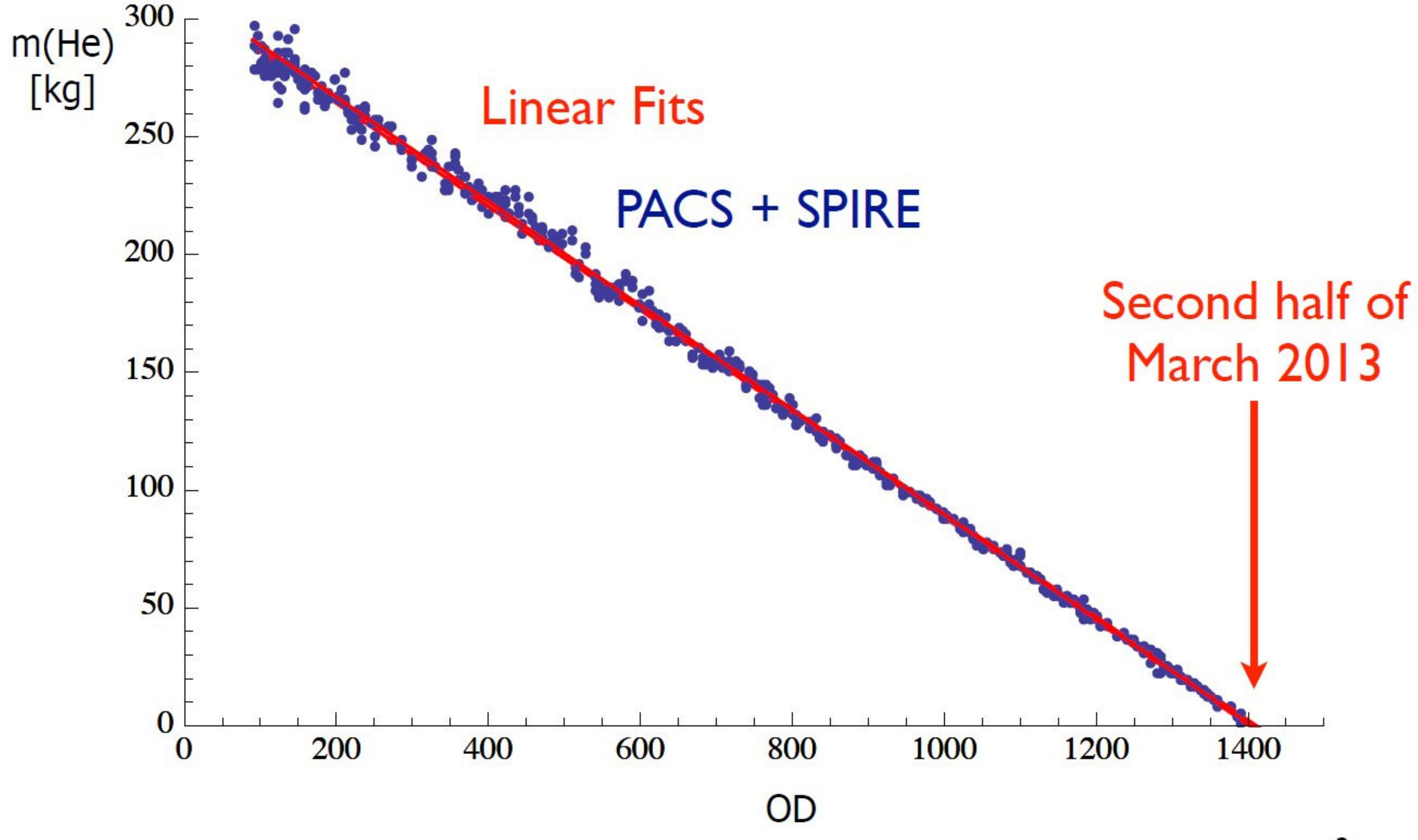}
  \end{center}
\caption{Top: Measured temperature rise of the superfluid $^4$He
L0 bath following the heat dissipation from the PACS and SPIRE cooler
recyclings. Red and blue dots correspond to the measurement with
two different temperature sensors. Bottom: End of Life (EOL) 
prediction of the superfluid $^4$He content in the {\it Herschel} cryostat.  
}
\label{fig:cooler-LHelifetime}       
\end{figure}

The PACS team cryogenic experts developed an alternative method
utilizing the regularly executed cooler recyclings
of PACS and SPIRE. During the recyclings a well
reproducible heat dissipation from the cooler pump into
the superfluid $^4$He L0 bath of the Herschel cryostat
took place. Fig.~\ref{fig:cooler-LHelifetime} (top) shows
the corresponding measured temperature rise, which is the higher
the less superfluid $^4$He remains. A smooth and well sampled
trend was observed with this method.

With the knowledge of the tank volume, the specific heat capacities 
of superfluid $^4$He and He gas in the ullage and the density
of both phases, the remaining superfluid $^4$He mass can be 
determined. This is plotted in Fig.~\ref{fig:cooler-LHelifetime} (bottom)
and led to a prediction for End of Life in the second half of 
March 2013. The actual superfluid $^4$He boil-off happened
on 29 April 2013, i.e. the superfluid $^4$He lifetime turned
out to be one month longer than predicted. This is owed to
uncertainties in both the adopted physical properties, like e.g.\
the values of the specific heat capacities, and the start conditions, 
namely the initial superfluid $^4$He mass in the tank at launch.

\section{The Slightly Non-nominal Cooler Recycling on OD\,1440}
\label{sec:nonnominal_recycling}

The shape of the relevant cooler temperatures during the recycling 
as shown in Fig.~\ref{fig:recycling_temperatures} was highly reproducible
during the whole {\it Herschel} mission until the last recycling
on OD\,1443, with one exception for the second-to-last recycling 
on OD\,1440, 1 week before the boil-off of the liquid $^4$He
in the large {\it Herschel} cryostat tank.  
  
\begin{figure}[h!]
  \begin{center}
  \includegraphics[width=1.0\textwidth]{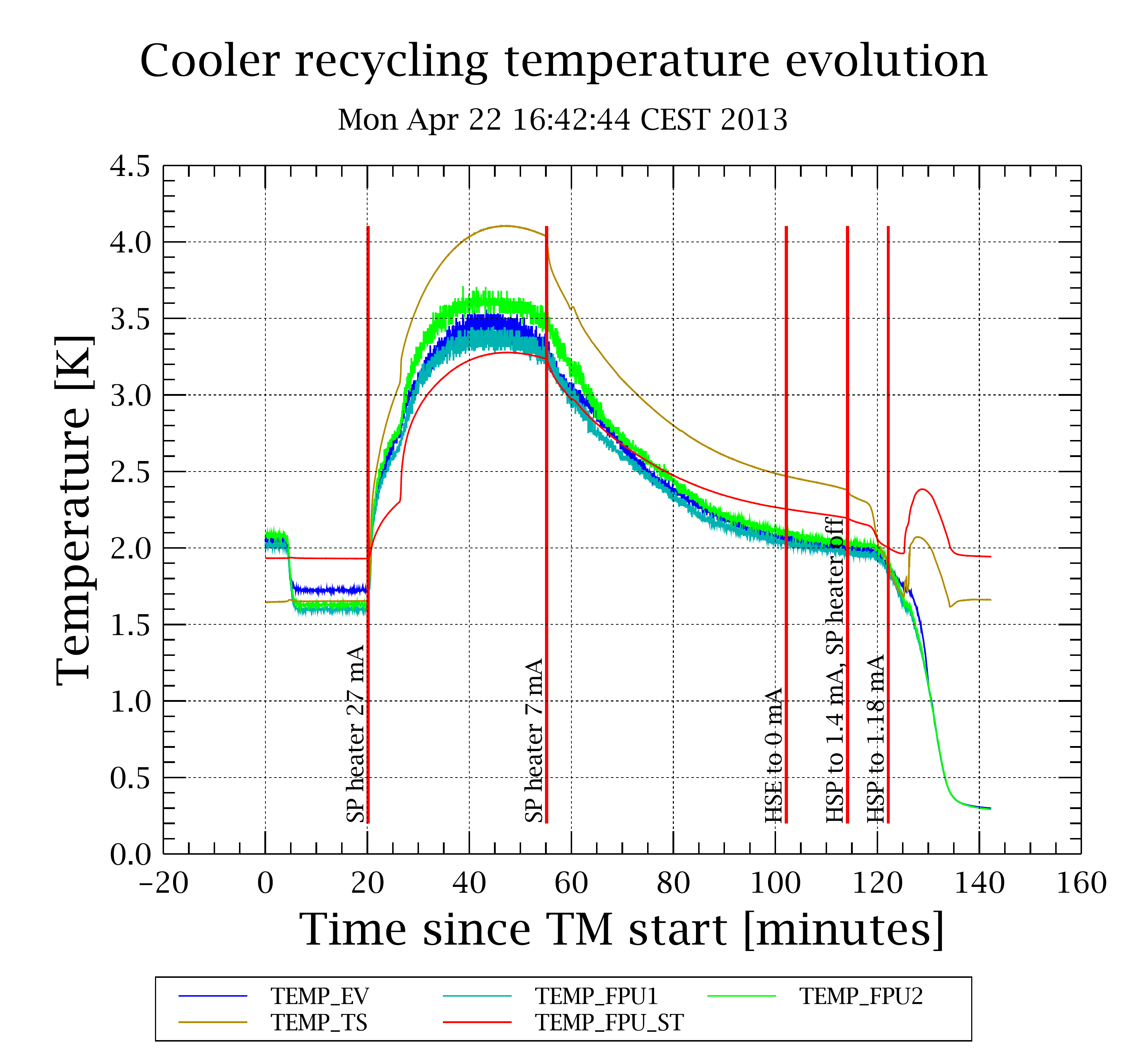}
  \end{center}
\caption{Evolution of temperatures relevant for the PACS cooler
monitored by sensors and provided in the PACS instrument
Housekeeping (HK) for the only slightly non-nominal recycling
on OD\,1440. 
}
\label{fig:recycling_nonnominal}       
\end{figure}

Fig.~\ref{fig:recycling_nonnominal} shows the temperature evolution
which can be compared with the one shown in Fig.~\ref{fig:recycling_temperatures}.
It can be noted that TEMP\_EV exceeds 3.5\,K during the heating of the 
pump and that the ``knee'' at around 117\,min after thermally disconnecting 
the evaporator (HSE = 0) and closing the pump heater switch (HSP)
is at 2\,K and not below. As pointed out in Sect.~\ref{sec:cooler_recycling},
the finally achieved TEMP\_EV in this step characterizes the efficiency 
of the recycling. This led to a 3.3\,h shorter cooler hold time
than anticipated (see Sect.~\ref{sec:holdtime_relation} and 
Table~\ref{tab:cooler_statistics_1}). Due to a safety buffer of 3\,h 
taken into account by the science mission planning only 0.3\,h of 
PACS photometer observations were lost\footnote{The earlier increase 
of the evaporator temperature triggered an autonomy function and put 
PACS into safe mode leading to the failure of OBSIDs 1342270750 \& 1342270751 
planned to observe asteroid (2000) Herschel.}.

The likely explanation for this behavior is that the remaining
liquid $^4$He film was broken or too thin so that the heat
dissipation from the pump was not perfect. This is also reflected
in the otherwise not observed bumps in TEMP\_TS and TEMP\_FPU\_ST
at about 128\,min. This event was the only sign before the final
boil-off, that the liquid $^4$He reserve was close to its end.
The liquid $^4$He film recovered again for the next and final PACS
cooler recycling on OD\,1443.

\section{Conclusion}
\label{sec:conclusion}

\begin{itemize}
\item[1)] The PACS $^3$He sorption cooler exceeded the required
          cooler hold time of 46\,h by at least 15\,h
          depending on the operational time of the PACS bolometer 
          detectors. The biased time of the detectors
          is the essential parameter for the resulting hold time.
          We did not observe any sign of aging effect of the device, 
          despite more than 9 years of operations, almost 4 of them in space.
\item[2)] An automatic cooler recycling procedure, 
          assembling the necessary steps with the right timing,
          worked highly reproducible over the whole mission
          until the last cooler recycling 4 days before
          the liquid $^4$He boil-off in the {\it Herschel}
          cryostat. The only slightly non-nominal recycling
          with a 5\% shorter hold time on OD\,1440 can be
          explained by a disturbed thermal interface to the
          liquid $^4$He bath 1 week before the boil-off.
          In total, 139 automatic PACS cooler recyclings
          and 100 automatic parallel (with SPIRE) cooler
          recyclings were executed during the Herschel mission. 
\item[3)] The dependence of the cooler hold time on the
          detector operational time (biased detectors)
          was established early in the mission and worked 
          reliably for the science mission planning process 
          over the entire {\it Herschel} mission.
\item[4)] We give a statistics of all PACS cooler recyclings
          over the full {\it Herschel} mission. From this
          fine differences in the performance of the
          different modes and start conditions can be seen.
          Parallel cooler recyclings starting with a warm
          cooler give an about 1\,h shorter hold time 
          than PACS-only recyclings. Cooler recyclings
          starting from a still cold cooler give an about
          0.5\,h longer hold time than the warm start 
          PACS-only recycling.
\item[5)] We characterize the cooler cycles into two
          main classes: (1) Dip-type cycles starting
          at a slightly higher level ($\approx$1\,mK 
          in evaporator temperature) and, for the majority,
          (2) swell-type cycles starting at a slightly 
          lower level and usually showing a temperature 
          swell of 1\,mK after around 24\,h adjusting then
          to the level of the dip-type cycles. The 
          physical reason for this ``swell'' is unclear, 
          it appears to be a zero gravity effect.
          The ``dip-type'' shapes were also observed
          during ground tests of the device and seem
          to be the consequence of relative timing
          of evaporator temperature evolution and 
          detector bias switching. We also noted a 
          systematically slightly higher evaporator 
          temperature level for parallel cooler cycles 
          during the time interval of SPIRE operations. 
          Illumination conditions of the detectors could 
          be ruled out as negligible. 
\item[6)] We investigated the cause for the dispersion
          of cooler hold times of cycles with similar 
          biased time, same mode and start condition,
          which is in the order of about 1\,h.
          By determining a zero bias hold time,
          allowing a homogeneous comparison of all
          cycles, we have seen that the dispersion
          is caused by trends and different thermal 
          loads into the PACS instrument along the
          {\it Herschel} mission. This was concluded
          from the fact that the PACS optical bench
          temperature sensor showed a quite good anti-correlation
          both with regard to long term trends and short term
          scatter. A few outliers with noticeable shorter hold
          times can be related to a confined period
          between OD\,101 and OD\,132 early in the mission,
          when larger temperature load variations occurred.
\item[7)] The evolution of the cooler evaporator temperature
          during the individual cooler cycles was studied
          and the impact of the temperature variation on the
          bolometer response and hence the photometric accuracy
          during the biased times could be characterized very 
          well and corrections improving the relative 
          calibration accuracy were established as described
          in \cite{Refmoor13}.
\item[8)] We sketch a method of $^4$He mass content determination
          evaluating the temperature response of the $^4$He
          coolant in the large {\it Herschel} cryostat to the
          constant heat deposition of the $^3$He sorption
          cooler pump at the end of the recycling process.
          This gave a much denser and hence robust against 
          individual outliers coverage of the mass content 
          curve for zero cost as an alternative to few dedicated 
          Direct Liquid Content Measurements (DLCM) requiring the 
          interrupt of the scientific operations. The accuracy
          of the method utilizing the anyhow performed cooler 
          recyclings is solely limited by uncertainties in the 
          adopted physical properties, like e.g.\ the values of 
          the specific heat capacities, and the start conditions, 
          namely the initial superfluid $^4$He mass in the tank 
          at launch.    
\end{itemize}

%
%
%
%
%
\begin{acknowledgements}
PACS has been developed by a consortium of institutes
led by MPE (Germany) and including UVIE (Austria);
KUL Leuven, CSL, IMEC (Belgium); CEA, LAM (France);
MPIA (Germany); INAF-IFSI/OAA/OAP/OAT, LENS, SISSA
(Italy); IAC (Spain). This development and the
operation of the PACS Instrument Control Centre,
planning the calibration observations, analyzing
and documenting the measurements and supporting
the mission planning stray-light protection has been
supported by the funding agencies BMVIT (Austria),
ESA-PRODEX (Belgium), CEA/CNES (France), DLR (Germany),
ASI/INAF (Italy), and CICYT/MCYT (Spain).
\end{acknowledgements}


\begin{thebibliography}{}
%
%
\bibitem{Refbalog13}
Balog Z.,~et al., The {\it Herschel}-PACS photometer calibration. Point-source flux calibration for scan maps, 
Exp Astron, DOI 10.1007/s10686-013-9352-3 this volume (2013)
\bibitem{Refbillot10}
Billot, N.~et al., CEA bolometer arrays: the first year in space, Proc.\ SPIE,
Vol.\ 7741, article id.\ 774102, 11 pp. (2010)
\bibitem{RefDuband95}
Duband, L.,~et al., {\it A thermal switch for use at liquid helium temperature in space-borne cryogenics systems}, Proceedings of the 8$^{th}$ International Cryocooler Conference, 1995, Plenum Press, NY, p. 731-741
\bibitem{Refduband08}
Duband L.,~et al., {\it Herschel} flight models sorption coolers, Cryogenics,
48, 95 (2008)
\bibitem{Refgriffin10}
Griffin M.J.~et al., The {\it Herschel}-SPIRE instrument and its in-flight performance, A\&A, 518, L3 (2010)
\bibitem{Refmoor13}
Mo{\'o}r A.~et al., PACS photometer calibration block analysis, Exp Astron, this volume (2013)
\bibitem{Refnielbock13}
Nielbock M.~et al., The {\it Herschel} PACS photometer calibration. A time dependent flux calibration for the 
PACS chopped point-source photometry AOT mode, Exp Astron, DOI 10.1007/s10686-013-9348-z this volume (2013)
\bibitem{Refpilbratt10}
Pilbratt G.L.~et al., {\it Herschel} Space Observatory, A\&A, 518, L1 (2010)
\bibitem{Refpoglitsch10}
Poglitsch A.~et al., The Photodetector Array Camera and Spectrometer (PACS) on the {\it Herschel} Space Observatory, A\&A, 518, L2 (2010)
\end{thebibliography}


\clearpage

\begin{landscape}

\begin{table}[h!]
\caption{PACS cooler statistics. Column 1 gives a running number,
  distinguishing between PACS only, labeled ``A'', and parallel, labeled
  ``B'', cooler recyclings. Column 2 lists the {\it Herschel}
 Operational Day (OD), when the cooler recycling took place and the
 columns 3 and 4 the start and end times indicated by yyyy-mm-ddThh:mm:ss 
 universal time. Column 5 indicates the start condition of the cooler
 recycling,  whether it started from a warm (w) cooler, with all liquid 
 $^3$He exhausted, or from a cold (c) cooler, with a remainder of liquid 
 $^3$He. Column 6 gives the resulting hold time; times flagged by $^{t}$ 
 are not the fully achievable hold times of this cycle, but are truncated 
 due to an early cold (c) start of a new recycling. These recyclings are 
 not used in figure~\ref{fig:holdtime_mission}. Columns 7 and 8 provide 
 information on the bolometer operations, giving the number of bias 
 periods and the total biased time per cooler cycle. Column 9 flags
 features of the temperature evolution of the cycle, s: swell 
 $\approx$24\,h after start of recycling, an additional $<$/$>$ indicates
 significantly shorter/longer than $\approx$24\,h, d: initial dip (see 
 Sect.~\ref{sec:statistics} for more details). Particular or 
 non-nominal cooler recyclings are flagged with an explanatory footnote.
}
\label{tab:cooler_statistics_1}       
\begin{tabular}{lrccclcrc}
\hline
\#   &  OD  & start cooler  & end cooler    &start& hold time &\# bias& biased time & shape \\
     &      & recycling (UT)& recycling (UT)&cond.&    (h)    &periods&    (h)      & T$_{EV}$\\  
\hline
A001 &   26 & 2009-06-08T22:40:14 & 2009-06-09T01:02:36 & w &  23.8043$^{t}$ &        1      &   1.1781    & s \\
A002 &   27 & 2009-06-10T00:44:57 & 2009-06-10T03:07:19 & c &  70.6215      &        1      &   6.6511    & - \\
A003 &   31 & 2009-06-14T06:32:58 & 2009-06-14T08:55:20 & w &  67.7156      &        1      &  21.1211    & s \\
A004 &   38 & 2009-06-21T07:50:34 & 2009-06-21T10:12:56 & w &  34.9333$^{t}$ &        1      &   2.5333    & - \\
A005 &   40 & 2009-06-22T21:03:09 & 2009-06-22T23:25:31 & c &  71.2997      &        2      &  15.5522    & s \\
A006 &   46 & 2009-06-28T17:33:17 & 2009-06-28T19:55:39 & w &  71.8910      &        1      &   6.4761    & s \\
A007 &   53 & 2009-07-06T02:56:06 & 2009-07-06T05:18:28 & w &  72.5103      &        1      &   2.5250    & d \\
A008 &   64 & 2009-07-16T14:47:08 & 2009-07-16T17:09:30 & w &  67.0481      &        2      &  26.5164    & s \\
A009 &   67 & 2009-07-19T14:46:13 & 2009-07-19T17:08:35 & w &  67.9420      &        1      &  24.6364    & s \\
A010$^a$&72 & 2009-07-24T14:43:23 & 2009-07-24T17:05:45 & w &     -         &        2      &  38.4600    & s \\
A011 &   84 & 2009-08-05T14:52:20 & 2009-08-05T17:14:42 & w &  62.9721      &        3      &  45.6658    & s \\
A012 &   92 & 2009-08-13T20:21:42 & 2009-08-13T22:44:04 & w &  68.0481      &        3      &  24.2375    & s \\
A013 &   96 & 2009-08-17T15:15:43 & 2009-08-17T17:38:05 & w &  65.4114      &        2      &  40.1767    & s \\
A014 &  101 & 2009-08-22T13:27:38 & 2009-08-22T15:50:00 & w &  67.0301      &        1      &  20.0683    & 2s \\
A015 &  104 & 2009-08-25T13:35:37 & 2009-08-25T15:57:59 & w &  64.8066      &        2      &  43.3583    & d \\
A016 &  107 & 2009-08-28T13:44:13 & 2009-08-28T16:06:35 & w &  45.8214$^{t}$ &        2      &  41.4394    & s$<$\\
A017 &  109 & 2009-08-30T13:50:13 & 2009-08-30T16:12:35 & c &  42.6895$^{t}$ &        2      &  37.7328    & s$<$\\
B001 &  111 & 2009-09-01T10:19:08 & 2009-09-01T13:11:40 & c &  61.5438      &        2      &  43.2303    & s \\
A018 &  118 & 2009-09-08T23:53:36 & 2009-09-09T02:15:58 & w &  58.1821$^{t}$ &        2      &  34.0006    & s$<$\\
A019 &  120 & 2009-09-11T12:21:16 & 2009-09-11T14:43:38 & c &  72.7466      &        1      &   5.2692    & s \\
A020 &  124 & 2009-09-14T23:42:21 & 2009-09-15T02:04:43 & w &  60.7143$^{t}$ &        2      &  43.5650    & s \\
A021 &  127 & 2009-09-17T14:42:05 & 2009-09-17T17:04:27 & c &  62.5602      &        2      &  43.7186    & s \\
A022 &  132 & 2009-09-23T01:45:58 & 2009-09-23T04:08:20 & w &  67.1199      &        1      &  18.3050    & d \\
B002 &  136 & 2009-09-26T23:12:08 & 2009-09-27T02:04:40 & w &  45.1275$^{t}$ &        2      &  37.5689    & d \\
\hline
\end{tabular} \\
$^a$ PACS switch-off during Onboard Software Upload in DTCP-74, no TM received
\end{table}

\clearpage

\addtocounter{table}{-1}

\begin{table}[h!]
\caption{PACS cooler statistics continued. 
}
\label{tab:cooler_statistics_2}       
\begin{tabular}{lrccclcrc}
\hline
\#   &  OD  & start cooler  & end cooler    &start& hold time &\# bias& biased time & shape \\
     &      & recycling (UT)& recycling (UT)&cond.&    (h)    &periods&    (h)      & T$_{EV}$\\  
\hline
A023 &  138 & 2009-09-28T23:06:36 & 2009-09-29T01:28:58 & c &  65.3623      &        2      &  40.3878    & d \\
A024 &  146 & 2009-10-06T22:43:56 & 2009-10-07T01:06:18 & w &  45.6248$^{t}$ &        1      &  20.6497    & d \\
A025 &  148 & 2009-10-08T22:38:06 & 2009-10-09T01:00:28 & c &  45.6031$^{t}$ &        2      &  37.1192    & s$>$\\
A026 &  150 & 2009-10-10T22:32:08 & 2009-10-11T00:54:30 & c &  67.0422      &        2      &  37.1500    & d \\
A027 &  156 & 2009-10-16T22:13:22 & 2009-10-17T00:35:44 & w &  64.7625      &        2      &  43.7536    & d \\
A028 &  159 & 2009-10-19T22:03:37 & 2009-10-20T00:25:59 & w &  21.6661$^{t}$ &        1      &  15.0633    & d \\ 
A029 &  160 & 2009-10-20T22:00:21 & 2009-10-21T00:22:43 & c &  46.0562$^{t}$ &        2      &  33.8381    & s$>$\\
B003 &  162 & 2009-10-22T21:53:52 & 2009-10-23T00:46:24 & c &  45.0942$^{t}$ &        4      &  40.4399    & s \\
A030 &  164 & 2009-10-24T21:47:26 & 2009-10-25T00:09:48 & c &  67.9650      &        2      &  20.6369    & 2s \\
A031 &  171 & 2009-10-31T21:25:59 & 2009-10-31T23:48:21 & w &  63.1170      &        3      &  48.5269    & s \\
A032 &  177 & 2009-11-06T21:08:55 & 2009-11-06T23:31:17 & w &  64.0964      &        3      &  42.9447    & 2s\\
A033 &  184 & 2009-11-13T20:49:39 & 2009-11-13T23:12:01 & w &  45.6627$^{t}$ &        2      &  35.5911    & d \\
A034 &  186 & 2009-11-15T20:46:04 & 2009-11-15T23:08:26 & c &  46.5388$^{t}$ &        2      &  40.0861    & d \\
A035 &  188 & 2009-11-17T21:36:12 & 2009-11-17T23:58:34 & c &  44.8176$^{t}$ &        1      &  19.9575    & s \\
A036 &  190 & 2009-11-19T20:42:49 & 2009-11-19T23:05:11 & c &  46.1684$^{t}$ &        2      &  33.3992    & 2s\\
B004 &  192 & 2009-11-21T20:41:22 & 2009-11-21T23:33:54 & c &  65.7532      &        2      &  35.2633    & s \\
A037 &  197 & 2009-11-26T20:38:22 & 2009-11-26T23:00:44 & w &  46.2087$^{t}$ &        2      &  41.7544    & s \\
B005 &  199 & 2009-11-28T20:37:28 & 2009-11-28T23:30:00 & c &  67.7568      &        2      &  31.2525    & d \\
B006 &  203 & 2009-12-02T20:36:11 & 2009-12-02T23:28:43 & w &  63.3581      &        3      &  45.2086    & s$<$ \\
A038 &  206 & 2009-12-05T20:35:46 & 2009-12-05T22:58:08 & w &  63.8295      &        3      &  45.5047    & s \\
A039 &  212 & 2009-12-11T17:42:52 & 2009-12-11T20:05:14 & w &  45.8827$^{t}$ &        2      &  42.0986    & s \\
B007 &  214 & 2009-12-13T17:22:18 & 2009-12-13T20:14:50 & c &  65.2871      &        2      &  42.7278    & s$<$ \\
A040 &  218 & 2009-12-17T17:24:38 & 2009-12-17T19:47:00 & w &  45.7505$^{t}$ &        2      &  39.1311    & s \\
A041 &  220 & 2009-12-19T17:26:16 & 2009-12-19T19:48:38 & c &  64.1315      &        3      &  48.4658    & d \\
A042 &  223 & 2009-12-22T17:21:24 & 2009-12-22T19:43:46 & w &  62.6322      &        3      &  53.5508    & d \\
B008 &  228 & 2009-12-27T17:35:37 & 2009-12-27T20:28:09 & w &  62.3849      &        3      &  48.6158    & s \\
A043 &  233 & 2010-01-01T17:49:03 & 2010-01-01T20:11:25 & w &  62.7316      &        3      &  49.7739    & s$<$\\
B009 &  236 & 2010-01-04T17:49:32 & 2010-01-04T20:42:04 & w &  62.6127      &        3      &  49.4731    & s$<$\\
B010 &  239 & 2010-01-07T17:55:47 & 2010-01-07T20:48:19 & w &  45.2951$^{t}$ &        2      &  31.5111    & s \\
A044 &  241 & 2010-01-09T18:00:14 & 2010-01-09T20:22:36 & c &  48.1649$^{t}$ &        2      &  39.6519    & s$<$\\
A045 &  243 & 2010-01-11T20:27:54 & 2010-01-11T22:50:16 & c &  63.9023      &        3      &  48.8775    & d \\
\hline
\end{tabular}
\end{table}

\clearpage

\addtocounter{table}{-1}

\begin{table}[h!]
\caption{PACS cooler statistics continued. 
}
\label{tab:cooler_statistics_3}       
\begin{tabular}{lrccclcrc}
\hline
\#   &  OD  & start cooler  & end cooler    &start& hold time &\# bias& biased time & shape \\
     &      & recycling (UT)& recycling (UT)&cond.&    (h)    &periods&    (h)      & T$_{EV}$\\  
\hline
A046 &  246 & 2010-01-14T20:35:12 & 2010-01-14T22:57:34 & w &  63.1364      &        3      &  48.3519    & s \\
A047 &  251 & 2010-01-19T18:25:06 & 2010-01-19T20:47:28 & w &  62.4450      &        3      &  49.9144    & d \\
B011 &  256 & 2010-01-24T18:38:36 & 2010-01-24T21:31:08 & w &  45.4533$^{t}$ &        2      &  38.9014    & s \\
A048 &  258 & 2010-01-26T18:52:33 & 2010-01-26T21:14:55 & c &  46.1565$^{t}$ &        2      &  42.9736    & d \\
B012 &  260 & 2010-01-28T18:49:38 & 2010-01-28T21:42:10 & c &  65.0416      &        3      &  39.5114    & d \\
B013 &  268 & 2010-02-05T19:11:52 & 2010-02-05T22:04:24 & w &  45.8134$^{t}$ &        2      &  34.1367    & d \\
B014 &  270 & 2010-02-07T19:17:24 & 2010-02-07T22:09:56 & c &  61.8602      &        3      &  54.8289    & s \\
B015 &  274 & 2010-02-11T20:35:08 & 2010-02-11T23:27:40 & w &  65.8847      &        2      &  28.8431    & s$<$\\
A049 &  284 & 2010-02-21T18:39:57 & 2010-02-21T21:02:19 & w &  61.7253      &        3      &  53.8569    & s$<$\\
A050 &  299 & 2010-03-08T22:40:03 & 2010-03-09T01:02:25 & w &  64.6146      &        2      &  39.6306    & s$<$\\
B016 &  314 & 2010-03-23T21:14:03 & 2010-03-24T00:06:35 & w &  60.7813      &        3      &  54.8858    & s \\
A051 &  320 & 2010-03-29T23:52:03 & 2010-03-30T02:14:25 & w &  62.9215      &        3      &  47.4594    & d \\
A052 &  324 & 2010-04-02T21:12:41 & 2010-04-02T23:35:03 & w &  64.2123      &        2      &  41.9975    & s \\
A053 &  328 & 2010-04-06T22:01:36 & 2010-04-07T00:23:58 & w &  63.6904      &        3      &  45.4486    & s \\
A054 &  343 & 2010-04-21T23:25:43 & 2010-04-22T01:48:05 & w &  62.4411      &        3      &  48.4106    & d \\
A055 &  346 & 2010-04-24T20:21:49 & 2010-04-24T22:44:11 & w &  62.2905      &        3      &  51.0150    & d \\
B017 &  349 & 2010-04-27T20:38:59 & 2010-04-27T23:31:31 & w &  45.6088$^{t}$ &        2      &  37.0225    & s \\
B018 &  351 & 2010-04-29T20:32:05 & 2010-04-29T23:24:37 & c &  45.3056$^{t}$ &        2      &  37.2692    & d \\
B019 &  353 & 2010-05-01T20:08:18 & 2010-05-01T23:00:50 & c &  65.1113      &        3      &  39.4364    & s$<$\\
A056 &  356 & 2010-05-04T20:15:04 & 2010-05-04T22:37:26 & w &  64.1593      &        2      &  43.4786    & s \\
A057 &  360 & 2010-05-08T19:55:45 & 2010-05-08T22:18:07 & w &  64.1401      &        2      &  43.0119    & d \\
B020 &  368 & 2010-05-16T19:42:22 & 2010-05-16T22:34:54 & w &  66.0664      &        2      &  27.3881    & s \\
A058 &  371 & 2010-05-19T19:37:37 & 2010-05-19T21:59:59 & w &  61.8683      &        3      &  53.6761    & s \\
B021 &  374 & 2010-05-22T19:32:55 & 2010-05-22T22:25:27 & w &  64.0344      &        3      &  39.7544    & s \\
B022 &  380 & 2010-05-28T18:56:03 & 2010-05-28T21:48:35 & w &  64.5570      &        2      &  38.3961    & s \\
A059 &  385 & 2010-06-02T18:40:01 & 2010-06-02T21:02:23 & w &  61.5922      &        3      &  55.7314    & s \\
B023 &  392 & 2010-06-09T14:47:42 & 2010-06-09T17:40:14 & w &  47.7938$^{t}$ &        2      &  28.9322    & s$<$\\
B024 &  394 & 2010-06-11T16:52:03 & 2010-06-11T19:44:35 & c &  64.7317      &        3      &  39.2725    & s$<$\\
A060 &  399 & 2010-06-16T15:59:20 & 2010-06-16T18:21:42 & w &  46.1451$^{t}$ &        2      &  43.7475    & d \\
B025 &  401 & 2010-06-18T15:54:41 & 2010-06-18T18:47:13 & c &  45.1666$^{t}$ &        2      &  36.1786    & s \\
B026 &  403 & 2010-06-20T15:22:12 & 2010-06-20T18:14:44 & c &  66.4402      &        3      &  33.4058    & s \\
\hline
\end{tabular}
\end{table}

\clearpage

\addtocounter{table}{-1}

\begin{table}[h!]
\caption{PACS cooler statistics continued. 
}
\label{tab:cooler_statistics_4}       
\begin{tabular}{lrccclcrc}
\hline
\#   &  OD  & start cooler  & end cooler    &start& hold time &\# bias& biased time & shape \\
     &      & recycling (UT)& recycling (UT)&cond.&    (h)    &periods&    (h)      & T$_{EV}$\\  
\hline
A061 &  413 & 2010-06-30T15:26:53 & 2010-06-30T17:49:15 & w &  61.5300      &        3      &  55.8081    & s$<$\\
B027 &  416 & 2010-07-03T15:19:57 & 2010-07-03T18:12:29 & w &  62.9263      &        3      &  45.7253    & s \\
B028$^b$&420& 2010-07-07T15:10:42 & 2010-07-07T18:03:14 & w &     -         &        2      &  37.7656    & s$<$\\
B029 &  434 & 2010-07-21T14:38:18 & 2010-07-21T17:30:50 & w &  45.1509$^{t}$ &        2      &  33.6669    & s \\
A062 &  436 & 2010-07-23T14:33:39 & 2010-07-23T16:56:01 & c &  64.8827      &        2      &  44.3497    & d \\
B030 &  446 & 2010-08-02T14:31:16 & 2010-08-02T17:23:48 & w &  45.6873$^{t}$ &        2      &  39.2767    & s \\
B031 &  448 & 2010-08-04T14:29:07 & 2010-08-04T17:21:39 & c &  45.6707$^{t}$ &        2      &  40.9700    & s \\
B032 &  450 & 2010-08-06T14:26:58 & 2010-08-06T17:19:30 & c &  45.2300$^{t}$ &        2      &  35.7411    & s \\
A063 &  452 & 2010-08-08T14:28:27 & 2010-08-08T16:50:49 & c &  62.1286      &        3      &  57.1875    & s$<$ \\
A064 &  455 & 2010-08-11T14:21:44 & 2010-08-11T16:44:06 & w &  61.5068      &        3      &  56.8969    & s \\
B033 &  458 & 2010-08-14T14:18:39 & 2010-08-14T17:11:11 & w &  61.6184      &        3      &  53.4636    & d \\
B034 &  464 & 2010-08-20T14:12:37 & 2010-08-20T17:05:09 & w &  63.8651      &        2      &  42.2944    & s \\
A065 &  469 & 2010-08-25T14:07:46 & 2010-08-25T16:30:08 & w &  61.5374      &        3      &  55.9964    & d \\
A066 &  472 & 2010-08-28T14:04:54 & 2010-08-28T16:27:16 & w &  62.0021      &        3      &  55.0281    & s \\
B035 &  478 & 2010-09-03T17:29:21 & 2010-09-03T20:21:53 & w &  45.6931$^{t}$ &        2      &  40.5706    & s \\
B036 &  480 & 2010-09-05T17:27:34 & 2010-09-05T20:20:06 & c &  63.5363      &        2      &  41.9306    & d \\
A067 &  483 & 2010-09-08T17:25:43 & 2010-09-08T19:48:05 & w &  62.3834      &        3      &  49.4347    & s \\
B037 &  486 & 2010-09-11T17:38:46 & 2010-09-11T20:31:18 & w &  64.1949      &        2      &  40.8369    & s \\
B038 &  492 & 2010-09-17T17:34:33 & 2010-09-17T20:27:05 & w &  63.7611      &        2      &  42.2675    & s \\
B039 &  497 & 2010-09-22T17:29:04 & 2010-09-22T20:01:36 & w &  45.6872$^{t}$ &        2      &  43.4728    & s \\
B040 &  499 & 2010-09-24T17:26:55 & 2010-09-24T20:19:27 & c &  44.9416$^{t}$ &        2      &  38.9772    & s \\
A068 &  501 & 2010-09-26T17:11:04 & 2010-09-26T19:33:26 & c &  64.9817      &        2      &  43.8539    & d \\
A069 &  511 & 2010-10-06T17:03:52 & 2010-10-06T19:26:14 & w &  61.4005      &        3      &  55.6553    & s \\
B041 &  514 & 2010-10-09T17:17:08 & 2010-10-09T20:09:40 & w &  70.8895      &        1      &  10.5675    & s$>$\\
B042 &  521 & 2010-10-16T16:57:42 & 2010-10-16T19:50:14 & w &  63.5320      &        2      &  39.9642    & s \\
B043 &  528 & 2010-10-23T16:53:24 & 2010-10-23T19:45:56 & w &  62.5161      &        3      &  45.6708    & s \\
A070 &  539 & 2010-11-03T16:12:19 & 2010-11-03T18:34:41 & w &  66.0888      &        2      &  34.3353    & s \\
A071 &  545 & 2010-11-09T16:13:54 & 2010-11-09T18:36:16 & w &  62.2343      &        3      &  50.3853    & s \\
A072 &  553 & 2010-11-17T16:05:23 & 2010-11-17T18:27:45 & w &  61.1577      &        3      &  57.5881    & s$<$\\
B044 &  564 & 2010-11-28T16:07:55 & 2010-11-28T19:00:27 & w &  61.1509      &        3      &  54.2536    & s$<$\\
\hline
\end{tabular} \\
$^b$ TM loss from OD\,422 to OD\,424
\end{table}

\clearpage

\addtocounter{table}{-1}

\begin{table}[h!]
\caption{PACS cooler statistics continued. 
}
\label{tab:cooler_statistics_5}       
\begin{tabular}{lrccclcrc}
\hline
\#   &  OD  & start cooler  & end cooler    &start& hold time &\# bias& biased time & shape \\
     &      & recycling (UT)& recycling (UT)&cond.&    (h)    &periods&    (h)      & T$_{EV}$\\  
\hline
B045 &  573 & 2010-12-07T12:53:27 & 2010-12-07T15:45:59 & w &  45.4319$^{t}$ &        2      &  40.8656    & s \\
B046 &  575 & 2010-12-09T12:35:57 & 2010-12-09T15:28:29 & c &  45.5978$^{t}$ &        2      &  41.8775    & d \\
B047 &  577 & 2010-12-11T12:29:33 & 2010-12-11T15:22:05 & c &  63.2060      &        3      &  50.3769    & d \\
A073 &  581 & 2010-12-15T12:13:43 & 2010-12-15T14:36:05 & w &  46.4597$^{t}$ &        2      &  44.6547    & d \\
B048 &  583 & 2010-12-17T12:27:57 & 2010-12-17T15:20:29 & c &  45.7077$^{t}$ &        2      &  43.1344    & d \\
B049 &  585 & 2010-12-19T12:28:04 & 2010-12-19T15:20:36 & c &  45.4999$^{t}$ &        2      &  40.1469    & s \\
B050 &  587 & 2010-12-21T12:15:46 & 2010-12-21T15:08:18 & c &  62.5957      &        3      &  51.6947    & d \\
B051 &  591 & 2010-12-25T12:24:44 & 2010-12-25T15:17:16 & w &  61.6580      &        3      &  53.7583    & s \\
B052 &  599 & 2011-01-02T12:02:37 & 2011-01-02T14:55:09 & w &  64.5291      &        2      &  39.0817    & s \\
B053 &  603 & 2011-01-06T10:44:26 & 2011-01-06T13:36:58 & w &  45.6712$^{t}$ &        2      &  39.1711    & d \\
B054 &  605 & 2011-01-08T10:41:29 & 2011-01-08T13:34:01 & c &  62.9152      &        3      &  49.5681    & s$<$ \\
A074 &  611 & 2011-01-14T10:18:37 & 2011-01-14T12:40:59 & w &  61.7411      &        3      &  54.5383    & s \\
A075 &  614 & 2011-01-17T10:13:08 & 2011-01-17T12:35:30 & w &  61.3934      &        3      &  57.7394    & s$<$\\
B055 &  619 & 2011-01-22T10:23:46 & 2011-01-22T13:16:18 & w &  61.4251      &        3      &  54.0169    & s \\
A076 &  627 & 2011-01-30T10:07:57 & 2011-01-30T12:30:19 & w &  60.9945      &        3      &  58.2303    & s \\
A077 &  635 & 2011-02-07T10:13:29 & 2011-02-07T12:35:51 & w &  62.0622      &        3      &  53.6019    & d \\
A078 &  638 & 2011-02-10T10:15:34 & 2011-02-10T12:37:56 & w &  61.6956      &        3      &  56.3108    & d \\
B056 &  647 & 2011-02-19T10:52:43 & 2011-02-19T13:45:15 & w &  61.6607      &        3      &  51.5978    & d \\
B057 &  651 & 2011-02-23T10:55:22 & 2011-02-23T13:47:54 & w &  61.8892      &        3      &  52.1847    & s \\
A079 &  661 & 2011-03-05T14:19:59 & 2011-03-05T16:42:21 & w &  61.2173      &        3      &  58.1769    & s \\
B058 &  668 & 2011-03-12T23:39:24 & 2011-03-13T02:31:56 & w &  61.1985      &        2      &  53.7753    & s \\
A080 &  675 & 2011-03-19T23:38:21 & 2011-03-20T02:00:43 & w &  61.2452      &        2      &  56.8564    & s \\
A081 &  684 & 2011-03-28T23:49:19 & 2011-03-29T02:11:41 & w &  60.8365      &        2      &  58.8703    & d \\
B059 &  695 & 2011-04-08T21:46:31 & 2011-04-09T00:39:03 & w &  61.0576      &        2      &  54.6689    & s \\
B060 &  701 & 2011-04-15T00:00:09 & 2011-04-15T02:52:41 & w &  61.3918      &        2      &  53.8089    & s \\
A082 &  704 & 2011-04-17T21:22:42 & 2011-04-17T23:45:04 & w &  61.2057      &        2      &  58.5483    & d \\
B061 &  712 & 2011-04-25T23:45:43 & 2011-04-26T02:38:15 & w &  45.7131$^{t}$ &        2      &  40.2983    & s \\
B062 &  714 & 2011-04-27T23:45:06 & 2011-04-28T02:37:38 & c &  62.5411      &        2      &  56.5261    & d \\
B063 &  718 & 2011-05-01T21:02:38 & 2011-05-01T23:55:10 & w &  61.7916      &        3      &  53.2950    & s \\
B064 &  721 & 2011-05-04T23:48:13 & 2011-05-05T02:40:45 & w &  62.2669      &        3      &  50.8091    & s \\
B065 &  724 & 2011-05-07T23:31:35 & 2011-05-08T02:24:07 & w &  61.6951      &        2      &  53.5144    & s \\
\hline
\end{tabular}
\end{table}

\clearpage

\addtocounter{table}{-1}

\begin{table}[h!]
\caption{PACS cooler statistics continued. 
}
\label{tab:cooler_statistics_6}       
\begin{tabular}{lrccclcrc}
\hline
\#   &  OD  & start cooler  & end cooler    &start& hold time &\# bias& biased time & shape \\
     &      & recycling (UT)& recycling (UT)&cond.&    (h)    &periods&    (h)      & T$_{EV}$\\  
\hline
B066 &  729 & 2011-05-12T23:21:24 & 2011-05-13T02:13:56 & w &  61.6653      &        2      &  53.3042    & d \\
A083 &  732 & 2011-05-15T23:02:01 & 2011-05-16T01:24:23 & w &  61.3188      &        2      &  58.7514    & d \\
B067 &  737 & 2011-05-20T23:09:35 & 2011-05-21T02:02:07 & w &  61.0989      &        2      &  56.1139    & s \\
A084 &  743 & 2011-05-26T22:46:20 & 2011-05-27T01:08:42 & w &  61.1842      &        1      &  58.7111    & s \\
B068 &  746 & 2011-05-29T22:56:30 & 2011-05-30T01:49:02 & w &  62.6281      &        3      &  48.8675    & s \\
B069 &  749 & 2011-06-01T22:51:35 & 2011-06-02T01:44:07 & w &  62.4941      &        2      &  52.5683    & s$<$ \\
A085 &  757 & 2011-06-09T18:44:50 & 2011-06-09T21:07:12 & w &  61.1722      &        1      &  58.5747    & s$<$\\
B070 &  762 & 2011-06-14T18:48:21 & 2011-06-14T21:40:53 & w &  60.8709      &        2      &  56.1197    & s \\
A086 &  769 & 2011-06-21T16:08:36 & 2011-06-21T18:30:58 & w &  60.9551      &        1      &  59.0025    & s \\
B071 &  775 & 2011-06-27T15:49:27 & 2011-06-27T18:41:59 & w &  61.9294      &        2      &  51.8719    & s \\
A087 &  780 & 2011-07-02T15:14:52 & 2011-07-02T17:37:14 & w &  61.1763      &        1      &  58.7919    & d \\
A088 &  787 & 2011-07-09T14:47:12 & 2011-07-09T17:09:34 & w &  61.5607      &        1      &  56.6278    & d \\
B072 &  790 & 2011-07-12T14:49:47 & 2011-07-12T17:42:19 & w &  61.2617      &        2      &  56.8269    & s \\
B073 &  793 & 2011-07-15T14:42:02 & 2011-07-15T17:34:34 & w &  61.1677      &        2      &  57.2194    & s \\
A089 &  797 & 2011-07-19T14:08:37 & 2011-07-19T16:30:59 & w &  61.6677      &        1      &  58.2642    & s \\
A090 &  800 & 2011-07-22T13:59:03 & 2011-07-22T16:21:25 & w &  61.5059      &        1      &  58.5600    & d \\
A091 &  805 & 2011-07-27T13:19:42 & 2011-07-27T15:42:04 & w &  61.2231      &        1      &  58.8314    & d \\
B074 &  808 & 2011-07-30T13:02:55 & 2011-07-30T15:55:27 & w &  61.8302      &        2      &  53.5567    & s \\
B075 &  824 & 2011-08-15T12:03:04 & 2011-08-15T14:55:36 & w &  63.5238      &        1      &  43.6217    & 2s\\
B076 &  828 & 2011-08-19T12:02:32 & 2011-08-19T14:55:04 & w &  60.4957      &        1      &  58.7331    & s \\
A092 &  831 & 2011-08-22T11:19:42 & 2011-08-22T13:42:04 & w &  61.0556      &        1      &  58.9658    & s \\
A093 &  842 & 2011-09-02T14:44:53 & 2011-09-02T17:07:15 & w &  60.7562      &        1      &  59.1175    & d \\
B077 &  847 & 2011-09-07T15:46:55 & 2011-09-07T18:39:27 & w &  60.4520      &        1      &  59.2228    & d \\
A094 &  858 & 2011-09-18T14:53:57 & 2011-09-18T17:16:19 & w &  60.9849      &        1      &  58.4603    & s \\
A095 &  871 & 2011-10-01T15:03:46 & 2011-10-01T17:26:08 & w &  60.8517      &        1      &  59.0897    & s \\
A096 &  887 & 2011-10-17T15:07:36 & 2011-10-17T17:29:58 & w &  60.8347      &        1      &  59.2342    & s \\
B078 &  892 & 2011-10-22T15:19:04 & 2011-10-22T18:11:36 & w &  61.3681      &        2      &  53.8206    & d \\
A097 &  898 & 2011-10-28T15:04:17 & 2011-10-28T17:26:39 & w &  61.0836      &        1      &  58.1575    & s$<$\\
B079 &  904 & 2011-11-03T15:16:11 & 2011-11-03T18:08:43 & w &  60.6198      &        1      &  57.3950    & s \\
B080 &  909 & 2011-11-08T15:54:23 & 2011-11-08T18:46:55 & w &  60.8049      &        2      &  56.6389    & s \\
A098 &  917 & 2011-11-16T15:01:17 & 2011-11-16T17:23:39 & w &  60.9123      &        1      &  58.4747    & d \\
\hline
\end{tabular}
\end{table}

\clearpage

\addtocounter{table}{-1}

\begin{table}[h!]
\caption{PACS cooler statistics continued. 
}
\label{tab:cooler_statistics_7}       
\begin{tabular}{lrccclcrc}
\hline
\#   &  OD  & start cooler  & end cooler    &start& hold time &\# bias& biased time & shape \\
     &      & recycling (UT)& recycling (UT)&cond.&    (h)    &periods&    (h)      & T$_{EV}$\\  
\hline
A099 &  926 & 2011-11-25T14:56:04 & 2011-11-25T17:18:26 & w &  60.9715      &        1      &  58.2014    & s \\
B081 &  930 & 2011-11-29T15:07:46 & 2011-11-29T18:00:18 & w &  61.4486      &        2      &  52.3406    & d \\
A100 &  934 & 2011-12-03T12:12:53 & 2011-12-03T14:35:15 & w &  61.1944      &        1      &  56.9378    & d \\
A101 &  945 & 2011-12-14T13:02:29 & 2011-12-14T15:24:51 & w &  60.8858      &        1      &  57.5639    & s \\
A102 &  955 & 2011-12-24T11:41:56 & 2011-12-24T14:04:18 & w &  60.9793      &        1      &  57.1031    & s \\
A103 &  967 & 2012-01-05T10:44:52 & 2012-01-05T13:07:14 & w &  60.8817      &        1      &  57.5353    & s \\
A104 &  973 & 2012-01-11T11:41:17 & 2012-01-11T14:03:39 & w &  61.0380      &        1      &  57.1114    & s \\
A105 &  981 & 2012-01-19T10:07:13 & 2012-01-19T12:29:35 & w &  61.0445      &        1      &  57.4642    & s \\
A106 & 1000 & 2012-02-07T10:36:17 & 2012-02-07T12:58:39 & w &  60.9549      &        1      &  57.3111    & s \\
B082 & 1005 & 2012-02-12T10:51:58 & 2012-02-12T13:44:30 & w &  60.4919      &        1      &  57.3442    & s \\
B083 & 1026 & 2012-03-04T13:41:39 & 2012-03-04T16:34:11 & w &  60.0322      &        1      &  57.3728    & s \\
A107 & 1034 & 2012-03-13T01:13:32 & 2012-03-13T03:35:54 & w &  61.4201      &        1      &  57.7531    & s \\
A108 & 1037 & 2012-03-15T23:22:28 & 2012-03-16T01:44:50 & w &  61.5980      &        1      &  57.6972    & d \\
A109 & 1040 & 2012-03-19T00:47:49 & 2012-03-19T03:10:11 & w &  61.5788      &        1      &  57.8047    & s \\
A110 & 1043 & 2012-03-22T00:13:45 & 2012-03-22T02:36:07 & w &  62.3290      &        3$^c$  &  52.5992    & d \\
A111 & 1049 & 2012-03-27T21:36:34 & 2012-03-27T23:58:56 & w &  61.2187      &        1      &  57.3392    & s \\
A112 & 1057 & 2012-04-04T21:36:15 & 2012-04-04T23:58:37 & w &  60.9807      &        1      &  57.3567    & s \\
B084 & 1063 & 2012-04-10T21:25:05 & 2012-04-11T00:17:37 & w &  62.0468      &        1      &  57.0419    & s \\
B085 & 1074 & 2012-04-21T21:10:53 & 2012-04-22T00:03:25 & w &  60.3257      &        1      &  57.4464    & s \\
B086 & 1081 & 2012-04-28T20:49:23 & 2012-04-28T23:41:55 & w &  60.5841      &        1      &  57.4281    & s \\
A113 & 1095 & 2012-05-12T19:48:01 & 2012-05-12T22:10:23 & w &  60.9746      &        1      &  57.6339    & d \\
B087 & 1101 & 2012-05-18T19:39:52 & 2012-05-18T22:32:24 & w &  60.2460      &        1      &  56.3936    & s \\
A114 & 1108 & 2012-05-25T19:01:50 & 2012-05-25T21:24:12 & w &  61.1698      &        1      &  57.5344    & s \\
A115 & 1119 & 2012-06-05T16:44:45 & 2012-06-05T19:07:07 & w &  61.0970      &        1      &  57.6556    & s \\
A116 & 1122 & 2012-06-08T16:43:08 & 2012-06-08T19:05:30 & w &  61.3202      &        1      &  57.7100    & d \\
A117 & 1136 & 2012-06-23T09:10:04 & 2012-06-23T11:32:26 & w &  60.8137      &        1      &  57.6911    & s \\
A118 & 1146 & 2012-07-02T15:57:44 & 2012-07-02T18:20:06 & w &  61.2490      &        1      &  57.1850    & s \\
A119 & 1157 & 2012-07-13T16:30:38 & 2012-07-13T18:53:00 & w &  60.9320      &        1      &  57.2533    & d \\
A120 & 1170 & 2012-07-26T14:40:38 & 2012-07-26T17:03:00 & w &  61.0454      &        1      &  57.6242    & s$<$ \\
B088 & 1179 & 2012-08-04T14:49:35 & 2012-08-04T17:42:07 & w &  60.4126      &        1      &  57.2525    & s$<$\\
\hline
\end{tabular} \\
$^c$ avoid interference with manual commanding (HIFI) 
\end{table}

\clearpage

\addtocounter{table}{-1}

\begin{table}[h!]
\caption{PACS cooler statistics continued. 
}
\label{tab:cooler_statistics_8}       
\begin{tabular}{lrccclcrc}
\hline
\#   &  OD  & start cooler  & end cooler    &start& hold time &\# bias& biased time & shape \\
     &      & recycling (UT)& recycling (UT)&cond.&    (h)    &periods&    (h)      & T$_{EV}$\\  
\hline
B089 & 1182 & 2012-08-07T15:17:36 & 2012-08-07T18:10:08 & w &  60.7152      &        1      &  56.8350    & s \\
A121 & 1193 & 2012-08-18T14:12:11 & 2012-08-18T16:34:33 & w &  61.0950      &        1      &  57.0972    & s \\
B090 & 1196 & 2012-08-21T13:52:24 & 2012-08-21T16:44:56 & w &  60.6987      &        1      &  57.2300    & d \\
B091 & 1200 & 2012-08-25T16:09:17 & 2012-08-25T19:01:49 & w &  60.8055      &        1      &  56.9783    & s$<$\\
B092 & 1214 & 2012-09-08T17:21:18 & 2012-09-08T20:13:50 & w &  60.1515      &        1      &  57.6483    & s \\
B093 & 1235 & 2012-09-29T18:36:12 & 2012-09-29T21:28:44 & w &  60.3889      &        1      &  56.3078    & s \\
A122 & 1244 & 2012-10-08T17:31:46 & 2012-10-08T19:54:08 & w &  61.0778      &        1      &  57.1506    & d \\
B094 & 1249 & 2012-10-13T16:54:47 & 2012-10-13T19:47:19 & w &  60.3564      &        1      &  57.4428    & s$<$ \\
B095 & 1264 & 2012-10-28T16:43:06 & 2012-10-28T19:35:38 & w &  60.4298      &        1      &  56.3814    & s$<$ \\
B096 & 1270 & 2012-11-03T16:40:49 & 2012-11-03T19:33:21 & w &  60.3439      &        1      &  57.5953    & d \\
B097 & 1273 & 2012-11-06T16:48:13 & 2012-11-06T19:40:45 & w &  60.4971      &        1      &  57.1886    & s \\
B098 & 1279 & 2012-11-12T21:45:51 & 2012-11-13T00:38:23 & w &  60.9443      &        1      &  54.7631    & s \\
A123 & 1285 & 2012-11-18T21:28:36 & 2012-11-18T23:50:58 & w &  60.8648      &        1      &  57.5872    & d \\
A124 & 1293 & 2012-11-26T21:24:49 & 2012-11-26T23:47:11 & w &  61.2999      &        1      &  54.7717    & s \\
A125 & 1308 & 2012-12-11T14:02:18 & 2012-12-11T16:24:40 & w &  60.7747      &        1      &  57.6894    & s \\
A126 & 1314 & 2012-12-18T02:36:52 & 2012-12-18T04:59:14 & w &  60.6358      &        1      &  57.8556    & d \\
A127 & 1320 & 2012-12-23T13:34:30 & 2012-12-23T15:56:52 & w &  61.0521      &        1      &  57.7108    & s \\
A128 & 1327 & 2012-12-30T13:26:59 & 2012-12-30T15:49:21 & w &  61.0103      &        1      &  57.4708    & s \\
A129 & 1332 & 2013-01-04T12:29:51 & 2013-01-04T14:52:13 & w &  60.8369      &        1      &  57.3789    & s$<$\\
A130 & 1337 & 2013-01-09T12:36:54 & 2013-01-09T14:59:16 & w &  60.9349      &        1      &  57.7964    & d \\
A131 & 1344 & 2013-01-16T13:01:53 & 2013-01-16T15:24:15 & w &  60.8892      &        1      &  57.4214    & d \\
A132 & 1349 & 2013-01-21T17:02:54 & 2013-01-21T19:25:16 & w &  60.7600      &        1      &  57.6244    & s \\
A133 & 1354 & 2013-01-26T17:32:26 & 2013-01-26T19:54:48 & w &  60.9113      &        1      &  57.7697    & d \\
B099 & 1375 & 2013-02-16T20:22:35 & 2013-02-16T23:15:07 & w &  60.8283      &        1      &  55.8261    & s \\
A134 & 1378 & 2013-02-19T23:33:12 & 2013-02-20T01:55:34 & w &  60.9005      &        1      &  57.4461    & s \\
A135 & 1399 & 2013-03-12T13:53:09 & 2013-03-12T16:15:31 & w &  60.3442      &        1      &  57.6919    & s$<$\\
B100 & 1402 & 2013-03-15T13:50:08 & 2013-03-15T16:42:40 & w &  60.4554      &        1      &  57.6394    & s \\
A136$^d$&1418& 2013-03-31T13:36:31& 2013-03-31T15:58:53 & w &  65.2737      &        1      &  56.3383    & d \\
A137 & 1426 & 2013-04-08T14:59:01 & 2013-04-08T17:21:23 & w &  60.8467      &        1      &  57.7586    & d \\
A138$^e$&1440& 2013-04-22T14:44:44& 2013-04-22T17:07:06 & w &  57.0020      &        1      &  57.7450    & s$<$ \\
A139 & 1443 & 2013-04-25T14:42:18 & 2013-04-25T17:04:40 & w &  61.0741      &        1      &  57.7806    & s \\
\hline
\end{tabular} \\
$^d$ not nominal: instrument power cycling in the middle of cooler cycle and biased time \\
$^e$ not nominal: anomaly in cooler recycling
\end{table}

\end{landscape}

\section{Appendix A: Common Uplink System (CUS) command script for PACS cooler recycling}
\label{sec:cooler_cus}

\begin{verbatim}
// Missionphase  : 
//
// Purpose       : Perform the cooler recycling
//
// TCL author    : TM
// TCL file      : tm_phot_cooler_recycling.tcl
// CUS author    : DAC
// Script file   : BOLO_cool_recycle.txt 
//
// Input arguments
// type    name         description
// N/A 
//
// Return values
// Type        Description
// int []      Several duration times
//
// Description   : see PhFPU UM, chapter 4
//
// Dependencies  : 
//
// Preconditions :
//
// Comments      : Based on the PhFPU UM Draft 6.0 and abundant e-mail
//                 exchanges with SAp
// 
// Version       : 2.0
//
int[] block BOLO_cool_recycle PACS 203 {
}{
    // disable AF 14 (to check for TEMP_EV < 0.3 K)
    Pacs_DPU_SET_FUNCT("EVENT_BOL_T_FPU","DISABLE");
    // disable AF 21 (to check if the absolute value of the HSP current is below 2*10^(-5)A)
    Pacs_DPU_SET_FUNCT("EVENT_BOL_I_SP1","DISABLE");
    // enable AF 18 (to check if the current of the sorption pump heater is below 30mA)
    Pacs_DPU_SET_FUNCT("EVENT_BOL_I_SP2","ENABLE");
    //
    // Obtain and set Block ID
    WriteBBID($BBID);
    // Define variables to communicate various durations to HSPOT.
    // NOTE: all time variables in units of number of ramps (SPEC) or
    //       number of readouts (BOLO). The calling program must convert this
    //       count into actual duration in true time units [seconds].
    //       SRC, REF, CAL, OVR stand for time spent on SRC, REF (on sky),
    //       CAL source, and overheads (wait for something). Total
    //       duration is given by duree_num. If no error, this duration
    //       must be equal to the sum of all others
    // NOTE: here all durations are given in [sec]
    int duree_num = 0;
    int duree_SRC = 0;
    int duree_REF = 0;
    int duree_CAL = 0;
    int duree_OVR = 0;
    // Set HK to PHOT
    Pacs_DPU_SET_HK_LIST("PHOT","BOTH Array");
    // content of "Preparation_au_recyclage.txt"
    // Set SP heater current to  to 0.00000000 amperes (0)
    // # P 07 01 0000
    int operand = 0x7010000;
    Pacs_DMC_SEND_COMMAND_BOLC(operand);
    int t_wait = 1;
    delay(t_wait);
    duree_num = t_wait;
    duree_OVR = t_wait;
    // Set HSP heater current to  to 0.00000000 amperes (0)
    // # P 07 02 0000
    operand = 0x7020000;
    Pacs_DMC_SEND_COMMAND_BOLC(operand);
    t_wait = 1;
    delay(t_wait);
    duree_num = duree_num + t_wait;
    duree_OVR = duree_OVR + t_wait;
    // Set HSE heater current to  to 0.00140000 amperes (3572)
    // # P 07 03 0DF4
    // Attendre 300000 ms
    // # S 01 0493E0
    operand = 0x7030df4;
    Pacs_DMC_SEND_COMMAND_BOLC(operand);
    t_wait = 300;
    delay(t_wait);
    duree_num = duree_num + t_wait;
    duree_OVR = duree_OVR + t_wait;
    // Set HSE heater current to  to 0.00118000 amperes (3011)
    // # P 07 03 0BC3
    // Attendre 900000 ms
    // # S 01 0DBBA0
    operand = 0x7030bc3;
    Pacs_DMC_SEND_COMMAND_BOLC(operand);
    t_wait = 900;
    delay(t_wait);
    duree_num = duree_num + t_wait;
    duree_OVR = duree_OVR + t_wait;
    //*****************************************************************
    //**                                                             **
    //**     Automatic Cooling reclycling                            **
    //**     ("Recyclage_Auto_Time.txt")                             **
    //**                                                             **
    //*****************************************************************
    //
    // - 23/01/06 Procedure of reclycling in Saclay cryostat with Phfpu MV
    // - note : this is timing version without temperature test
    //
    //         Initialisation of BOLC
    //
    // Set temp probe on/off FF hexa
    // # P 07 00 00 FF
    // 
    //         Initialisation of LTU
    //
    // Inhiber enregistrement TM
    // # S 08 
    // Valider enregistrement TM
    // # S 09
    //
    // Initial conditions
    // TEMP_SP < 10K
    // TEMP_EV < 2K
    //
    // Set SP heater current to  to 0.02730000 amperes (2231)
    // # P 07 01 08B7
    //
    // Attendre 2100000 ms
    // # S 01 200B20
    operand = 0x70108b7;
    Pacs_DMC_SEND_COMMAND_BOLC(operand);
    t_wait = 2100;
    delay(t_wait);
    duree_num = duree_num + t_wait;
    duree_OVR = duree_OVR + t_wait;
    //
    // Set SP heater current to  to 0.00700000 amperes (580)
    // # P 07 01 0244
    // 
    // Attendre 2820000 ms
    // # S 01 2B07A0
    operand = 0x7010244;
    Pacs_DMC_SEND_COMMAND_BOLC(operand);
    t_wait = 2820;
    delay(t_wait);
    duree_num = duree_num + t_wait;
    duree_OVR = duree_OVR + t_wait;
    //
    // Set HSE heater current to  to 0.00000000 amperes (0)
    // # P 07 03 0000 
    //
    // Attendre 720000 ms
    // # S 01 0AFC80 
    operand = 0x7030000;
    Pacs_DMC_SEND_COMMAND_BOLC(operand);
    t_wait = 720;
    delay(t_wait);
    duree_num = duree_num + t_wait;
    duree_OVR = duree_OVR + t_wait;
    //
    // Set SP heater current to  to 0.00000000 amperes (0)
    // # P 07 01 0000
    operand = 0x7010000;
    Pacs_DMC_SEND_COMMAND_BOLC(operand);
    t_wait = 1;
    delay(t_wait);
    duree_num = duree_num + t_wait;
    duree_OVR = duree_OVR + t_wait;
    //
    // Set HSP heater current to  to 0.00140000 amperes (3569)
    // # P 07 02 0DF1
    // 
    // Attendre 480000 ms
    // # S 01 075300 
    operand = 0x7020df1;
    Pacs_DMC_SEND_COMMAND_BOLC(operand);
    t_wait = 480;
    delay(t_wait);
    duree_num = duree_num + t_wait;
    duree_OVR = duree_OVR + t_wait;
    //
    // Set HSP heater current to  to 0.00118000 amperes (3010)
    // # P 07 02 0BC2 
    operand = 0x7020bc2;
    Pacs_DMC_SEND_COMMAND_BOLC(operand);
    t_wait = 1;
    delay(t_wait);
    duree_num = duree_num + t_wait;
    duree_OVR = duree_OVR + t_wait;
    //
    // ************************	
    // **                    **
    // ** end recycling      **
    // **                    **
    // ************************	
    //  Last message
    debug_print("Recycling completed, in about 20 min, TEMP_EV < 0.3 K");
    t_wait = 1200;
    duree_num = duree_num + t_wait;
    duree_OVR = duree_OVR + t_wait;
    delay(t_wait);
    //
    // AF settings
    //
    // disable AF 18 (to check if the current of the sorption pump heater is below 30mA)
    Pacs_DPU_SET_FUNCT("EVENT_BOL_I_SP2","DISABLE");
    // enable AF 21 (to check if the absolute value of the HSP current is below 2*10^(-5)A)
    Pacs_DPU_SET_FUNCT("EVENT_BOL_I_SP1","ENABLE");
    //
    // Set HK to NO PRIME (to have the clean SAFE settings again)
    Pacs_DPU_SET_HK_LIST("NO_PRIME","BOTH Array");
    // Return the array of times
    int[] time_array = [duree_num,duree_SRC,duree_REF,duree_CAL,duree_OVR];

    return time_array;
}
\end{verbatim}

\end{document}